# Large Language Model-Based Knowledge Graph System Construction for Sustainable Development Goals: An AI-Based Speculative Design Perspective

Lin, Yi-De; Liao, Guan-Ze

## 1. Abstract

2000-2015, the UN's Millennium Development Goals established eight targets guiding global resources. The subsequent Sustainable Development Goals (SDGs) adopted a more dynamic approach with annual indicator refinements. With 2030 approaching and limited progress, innovative acceleration strategies are urgently needed.

This study develops an AI-powered knowledge graph system to analyze SDG interconnections identify potential new goals and visualize them on the web. The dataset comprises official SDG documents, Elsevier's SDG keyword dataset, and TED Talk transcripts from 2020.01 to 2024.04 (1,127 talks), with a 2023 subset (269 talks) as a pilot study. The methodology combines AI-speculative design, large language models, and retrieval-augmented generation to construct a knowledge graph simulating SDG discussions.

Findings: (1) Heatmap analysis of SDG attribute tags in TED Talks indicates that data associated with Goal 10 and Goal 16 exhibits the highest proportions, suggesting a strong potential correlation. Conversely, Goal 6 shows the lowest proportion, implying a global deficiency in knowledge or discourse related to this goal. (2) In the knowledge graph of the system, as the duration of simulated discussions increases, a highly connected node distinct from the initial one emerges, indicating that in the context of AI-speculative design, richer and more relevant data enhances the identification of core concepts through divergent thinking, potentially improving goal execution. (3) The study identifies six potential new goals: inclusive well-being, poverty reduction through technological advancement, climate resilience for vulnerable communities, inclusive and equitable development, global collaboration for water security, and inclusive economic empowerment. These findings reinforce the centrality of Goal 1 and highlight the global emphasis on Goal 10.

This AI-speculative approach provides novel insights for enhancing the SDG framework, offering strategic guidance for policymakers while extending future applications to multimodal AI integration and cross-system synthesis for advancing global sustainability.

# 2. Introduction

## 2.1. Background and Motivation

2030 is the deadline for achieving the Sustainable Development Goals (SDGs). From 2000 to 2015, the Millennium Development Goals (MDGs) proposed eight goals with data-driven performance indicators, serving as a reference for the SDGs introduced in 2015. Despite regular expert assemblies to adjust indicators, actual progress has fallen short of expectations, leading scholars to question the SDGs framework's efficacy (Kim, 2023). Finding ways to promote sustainable global development is therefore urgently important.

Speculative Design creates space for exploring possibilities beyond reality's constraints, fostering dialogue while suggesting imaginative yet grounded possibilities. AI-speculative design extends this approach by using artificial intelligence to enhance imaginative communication, introduce external resources, and diversify algorithmic thinking. This allows AI to actively participate in virtual societal experiments that could support addressing SDG challenges.

Meanwhile, TED (Technology, Entertainment, and Design), established in 1984, aims to stimulate imagination and broaden knowledge. Evolving from "Ideas Worth Spreading" to "Ideas Change Everything," TED brings together experts and ideas from diverse fields, igniting imagination and inspiring transformative thinking.

The SDGs, AI-speculative design, and TED Talks share commitments to inclusivity, diverse participation, and scientific speculation about future developments. While existing research mainly uses AI to digitize SDG indicator tracking (Allen et al., 2024), limited work explores the SDG framework from an imaginative standpoint. By leveraging AI's reasoning and simulation abilities while synthesizing insights from TED Talks, AI could potentially develop a deeper understanding of sustainability issues than humans, fostering innovative thinking about the SDG framework.

## 2.2. Research Purposes and Questions

Under globalization, humanity is becoming increasingly interdependent while facing limited natural resources. The Sustainable Development Goals (SDGs), launched in 2015 with a 2030 deadline, have shown limited impact, suggesting the framework needs improvement (Kim, 2023).

This research aims to cultivate an imaginative environment for discussing a new sustainable goals framework through AI-speculative design, using TED Talks featuring diverse experts. The study develops a quasi-Socratic-style simulated conference system using large language models (LLMs), with three objectives:

1. Investigate data processing workflows integrating LLMs, Retrieval-Augmented Generation, and knowledge graph technologies

2. Develop an AI-speculative design environment simulating conference discussions on sustainable development

3. Examine the information structure of a conference-based knowledge graph using visualization techniques

The system personifies knowledge concepts as spokespeople, with SDGs as hosts and TED Talks data as participants. Through association mining on the SDGs knowledge graphs, the study aims to uncover potential new sustainable development goals that can supplement existing SDGs and guide future global development.

The dataset includes official SDG indicators, Elsevier's (2023) SDG keywords, and English TED Talk transcripts (4-20 minutes) from two periods: 2023 (269 videos) and 2020.01-2024.04 (1,127 videos). The latter period covers significant challenges like the COVID-19 pandemic following the 2019 UN Global Sustainable Development Report. Research questions address:

1. Identifying SDG attribute trends and relationships in TED Talks from 2020.01-2024.04.

2. Examining knowledge graph construction through simulated conference discussions.

3. Revealing new SDG-related content and indicators by compiling knowledge graphs to uncover hidden relationships and develop strategies for accelerating SDG achievement.

# 3. Related Work

## 3.1 SDGs

### 3.1.1 Core Values of the SDGs: Establishing the Goals for Humanity and the Earth

The Sustainable Development Goals (SDGs) comprise 17 primary objectives, emphasizing economic, social, and environmental dimensions (Purvis et al., 2019). They align with the UN's mission to eradicate poverty, inequality, and address climate change (United Nations, n.d.-c). UNDESA oversees SDG progress, using detailed goals and indicators to solidify "sustainable development" (United Nations, n.d.-d).. Academic interest in SDGs is growing, with sustainability publications on the Web of Science increasing significantly (Yumnam et al., 2024). Platforms like Elsevier and Scopus enhance SDG research accessibility (James, 2021, 2022, 2023). SDG 17 highlights partnerships among governments, businesses, and communities (Leal Filho et al., 2024).

### 3.1.2 Changes and Evolution of SDGs: From the Expansion of MDGs to the Recent Indicators

SDGs succeeded the Millennium Development Goals (MDGs), which addressed peace, poverty, and human rights (United Nations, 2000, 2008). SDGs, launched in 2015, include 17 goals and 169 sub-goals, focusing on human rights, planet, prosperity, peace, and partnership (United Nations, 2015, 2018). The UN developed detailed SDG indicators in 2016 (United Nations, 2016; United Nations, n.d.-a). Indicators have been adjusted from 2017 to 2024 (United Nations, 2020, 2021, 2022, 2023). SDG indicator modifications follow strict procedures and official statistics principles (United Nations Economic Commission for Europe, n.d.; United Nations, 2014). SDG 16, 10, 12, and 17 have seen the most indicator changes (International Institute for Sustainable Development, n.d.-d; United Nations Statistics Division, n.d.-d). SDGs emphasize global cooperation more than MDGs (Servaes, 2017; Kumar et al., 2016; United Nations Statistics Division, n.d.-d). Data collection changes and events like the COVID-19 pandemic have impacted SDG progress (United Nations, 2020; United Nations Statistics Division, n.d.-d, n.d.-c).

**Figure 3-1:** Timeline of SDGs Content Revisions

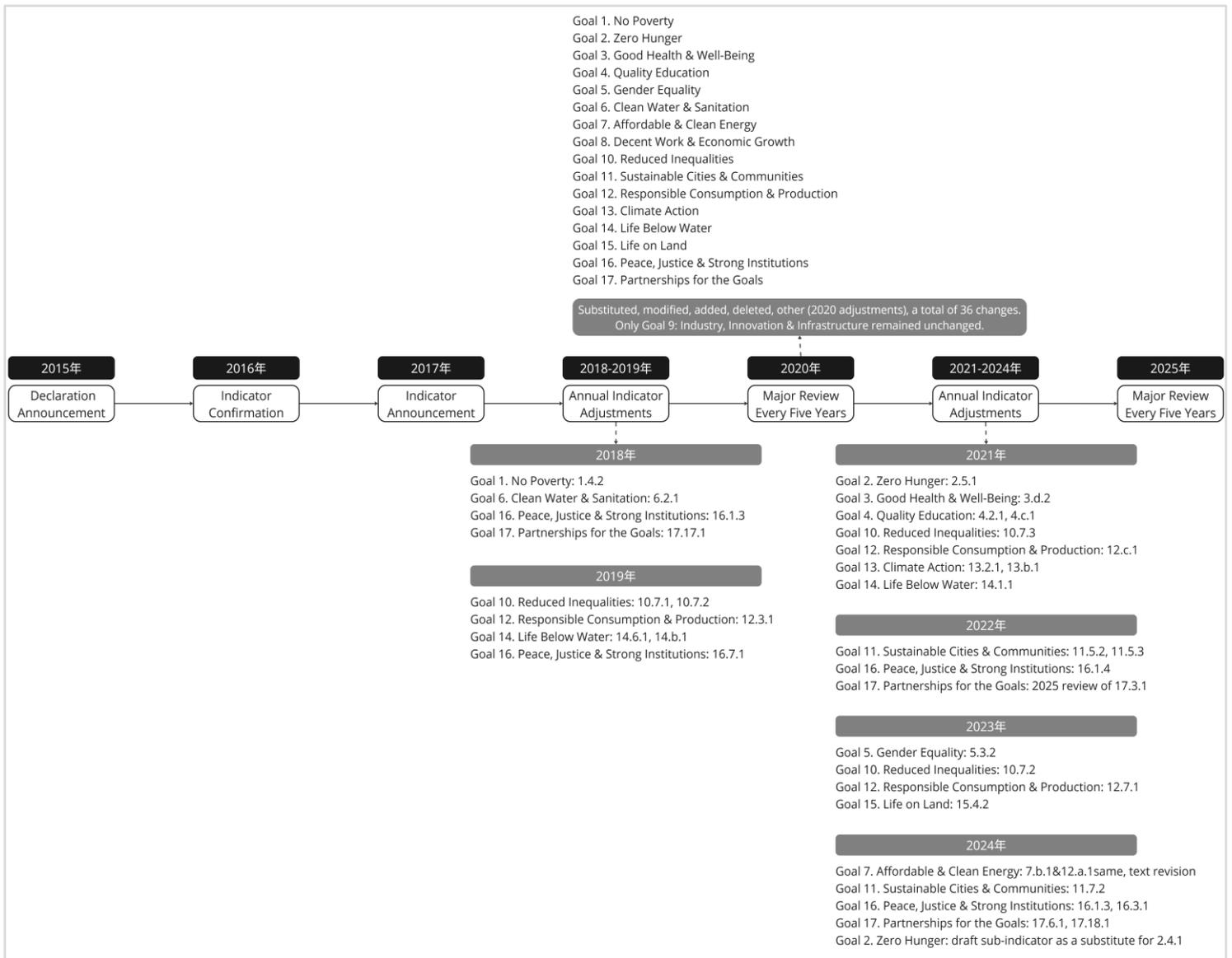

**Note:** Data compiled from various annual meeting documents. Meeting documents from *IAEG-SDGs — SDG Indicators*. (n.d.-c). https://unstats.un.org/sdgs/iaeg-sdgs/report-iaeg-sdgs

**Table 3-1**: Number of Changes in SDG Indicators from 2018 to 2024

| Goals | Number of Changes |
|---|---|
| Goal 1. No Poverty | 2 |
| Goal 2. Zero Hunger | 3 |
| Goal 3. Good Health & Well-Being | 2 |
| Goal 4. Quality Education | 2 |
| Goal 5. Gender Equality | 2 |
| Goal 6. Clean Water & Sanitation | 2 |
| Goal 7. Affordable & Clean Energy | 2 |
| Goal 8. Decent Work & Economic Growth | 1 |
| Goal 9. Industry, Innovation & Infrastructure | 0 |
| Goal 10. Reduced Inequalities | 4 |
| Goal 11. Sustainable Cities & Communities | 3 |
| Goal 12. Responsible Consumption & Production | 4 |
| Goal 13. Climate Action | 2 |
| Goal 14. Life Below Water | 3 |
| Goal 15. Life on Land | 2 |
| Goal 16. Peace, Justice & Strong Institutions | 5 |
| Goal 17. Partnerships for the Goals | 4 |

**Note:** Data compiled from various annual meeting documents. Meeting documents from *IAEG-SDGs — SDG Indicators*. (n.d.-c). https://unstats.un.org/sdgs/iaeg-sdgs/report-iaeg-sdgs/

## 3.1.3 Challenges Faced by the SDGs: Issues Arising from an Ambiguous Framework

SDGs face challenges due to framework ambiguity and limited contextual coverage. Despite increased indicators, they are deemed insufficient (Kim, 2023). Complex interrelations exist among SDGs, with some showing negative correlations (Nilsson et al., 2016; Pradhan et al., 2017). SDG understanding and application vary across countries (Mair et al., 2018; Tosun & Leininger, 2017). SDG implementation has been hindered, with many goals facing major challenges, and 2023 data shows stagnant progress in most sub-goals, with significant variations among regions (Sachs et al., 2023). These challenges highlight the need for broader discussion and reflection on the SDGs framework.

## 3.2. TED Talks

### 3.2.1 History, Core Values, and Philosophy of TED Talks: Expanding the Influence of Knowledge

TED Talks aim to promote diverse thinking, aligning with the need for broader reflection on the SDGs TED, n.d.-a). Chris Anderson revitalized TED in 2001, emphasizing "inspiring, broad, and diverse content" (Anderson, n.d., 2008). TED explores questions across fields, evolving from "Ideas Worth Spreading" to "Ideas Change Everything" (Anderson, 2024). TED's online talks, starting in 2006, feature diverse speakers. TED uses TEDx licensing and website updates to expand its reach (TED, n.d.-b; *(The new Ted.com lets you dig deeper...*, 2014). TED's core values—curiosity, reason, inclusivity, and impact—support knowledge sharing (TED, n.d.-c).

### 3.2.2 The Structural Features of TED Talks: Effectively Conveying Ideas

TED Talks focus on connection, narrative, explanation, persuasion, and revelation, with a strict 18-minute limit (Anderson, 2016; *TED Help Center*, n.d.). Talks are rigorously edited, and the recent addition of AI subtitling and translation significantly enhances accessibility (TED Localization Team, 2024; TED, n.d.-d). Anderson highlights the importance of a core idea, clear language, and audience engagement (Anderson, 2016). Scholars note TED's role in popularizing knowledge, using accessible language and storytelling (Mattiello, 2017; Sugimoto & Thelwall, 2013).

**3.2.3 TED Talks Inspire Imagination and Impact Sustainability Issues**

TED features a sustainability category, reflecting its impact on social awareness (TED, n.d.-e). Scholars explore TED's role in advocating for "environmental rights" and analyzing livable cities (Anesa, 2018; Yılmaz & Atay, 2023). TED inspires real-world impact through its "Ideas Change Everything" motto (Anderson, 2024). Research shows TED fosters innovation, influenced by speaker delivery and content relatability (Wang, 2021). TED Talks offer a flexible knowledge base for SDG discussions, promoting innovative approaches through speculative design.

## 3.3 Speculative Design
### 3.3.1 The Value of Speculative Design: Creating a Space for Imagination

Speculative Design, introduced by Dunne and Raby, aims to inspire creativity and provoke thinking about complex global issues (Dunne et al., 2019). It relates to Critical Design, which reflects on technology, art, and identity, driving change (Dunne, 2008). Material Speculation combines these, using objects to explore human-object-technology relationships (Wakkary et al., 2015). Speculative Design operates between the possible and imaginary, exploring "plausible" and "probable" scenarios (Dunne et al., 2019). Researchers examine participant roles and "Gap" areas, emphasizing uncertainty and interdisciplinary interaction (Meskus & Tikka, 2024). Effective speculative designs use "cognitive estrangement," like *Black Mirror*, to prompt reflection (Dunne et al., 2019). Studies use counterfactual approaches and workshops to envision future artifacts (Forlano & Halpern, 2023).

**Figure 3-2:** The scope of discussion in speculative design

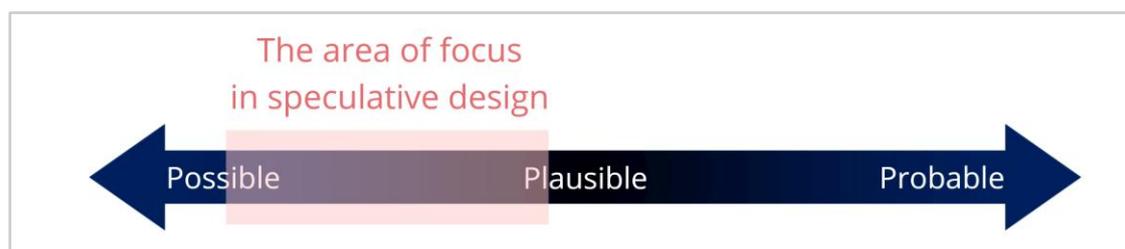

**Note**: From the past, present, to the future, every moment experiences various types of possibilities. If we cross-section a single point in time to examine these possibilities, they can be divided from left to right into Possible, Plausible, and Probable. Speculative design focuses on the section between Plausible and Probable (marked in pink). Adapted from the book: Dunne, Anthony, Hong Shih-Min, and Fiona Raby. *Speculative Design: Design, Imagination, and Social Dreams*. 1st Edition. Taipei: Ho Chiao-Wei Studio Publishing, 2019. Print.

**3.3.2 Speculative Design Applications Initiate Discussions on Sustainability Issues**

Speculative Design, like the box in "The Little Prince," evokes imagination (Mitrović et al., 2021). It lacks rigid norms, creating tension within design frameworks. Mitrović et al. (2021) suggest tension can arise from "differences" in exclusion, engagement, use, completion, gain, perspective, time, and inclusion. Speculative Design's tension drives sustainable development discussions, balancing social, economic, and environmental factors (Design for Sustainability Studio, n.d.). It fosters critical thinking among younger generations, using scenario simulations and role-playing (Angheloiu et al., 2020). Applications include community food systems and transportation, with workshops addressing climate change impacts (S. Chopra et al., 2022). Systemic frameworks, like urban transportation planning, use literature analysis to promote sustainability (Sustar et al., 2020). Speculative Design provides starting points, influenced by individual worldviews, fostering idea exchange and driving sustainable development.

## 3.4 Knowledge Graph
### 3.4.1 The Origin and Application Value of Knowledge Graphs

Knowledge Graphs and Ontologies are related, both representing and managing knowledge, though with distinctions (H. Chen & Luo, 2019; Silva et al., 2022). Knowledge Graphs model society as a system with interconnected entities. Ontologies, from philosophy, focus on systematic knowledge frameworks, unlike Knowledge Graphs' emphasis on graph structures (Gruber, 1995). Ontologies use semantic networks like OWL and RDF for structured knowledge bases (Gruber, 1995; Hitzler et al., 2009). Knowledge Graphs use nodes and links (edges) to depict knowledge as triples (Subject-Predicate-Object), forming dynamic semantic networks. Graphs can be directed or undirected, and enhanced through interconnections and external resources (Ehrlinger & Wöß, 2016; IBM Technology, 2022; Xiao, 2021). Since 2012, Google's search engine application has popularized Knowledge Graphs (Google, n.d.-a; Singhal, 2012). They're used in network node summarization, recommendation systems, business information management, and risk assessment (Bertram et al., 2023; Deng et al., 2022; The Knowledge Graph Conference, 2023).

## 3.4.2 Construction and Visualization of Knowledge Graphs: Temporal Knowledge Graphs and Tools

Knowledge Graph construction involves defining purpose, entity/relationship extraction, and data verification/maintenance (Ehrlinger & Wöß, 2016; Simsek et al., 2021; Xiao, 2021). Tools, including machine learning, aid in knowledge inference (X. Chen et al., 2020). Temporal Knowledge Graphs, for event-centered data, use timelines and time-based models (Knez & Žitnik, 2023). Applications include historical population contexts and elderly fall risk simulations (Gautam et al., 2020; Egami et al., 2023). A variety of visualization tools have emerged from semantic web technologies such as OWL and RDF, including JavaScript and Python libraries like D3.js, Cytoscape.js, Neo4j, NetworkX, and Plotly (Bizer, n.d.; World Wide Web Consortium [W3C], n.d.-a, n.d.-b, n.d.-c; Bostock, n.d.; Neo4j, 2020; NetworkX Developers, n.d.; Franz, n.d.).

## 3.4.3 Knowledge Graph Association Rule Mining: Discovering New Potential Directions within SDGs

Human thought uses various "representations" (Pinker, 2015). Knowledge Graphs structure knowledge into triples. Association rule mining explores relationships within these representations (Agrawal et al., 1993). Machine learning methods like inductive logic programming (ILP) and neural networks aid in this (H. Wu et al., 2023). Knowledge Graphs aid in knowledge management for sustainability, as seen in SustainGraph's SDG tracking (Erdl, 2023; Fotopoulou, Mandilara, Zafeiropoulos, Laspidou, et al., 2022; Fotopoulou, Mandilara, Zafeiropoulos, & Papavassiliou, 2022).

## 3.5 Large Language Models
### 3.5.1 Historical Development of Large Language Models

The 1990s saw increased human-computer interaction discussions. Deep learning and neural networks emerged in the 2000s, leading to pre-trained language models like Word2Vec in the 2010s. Recent breakthroughs with BERT and GPT have popularized generative AI and LLMs (BBC News 中文, 2019; Jana et al., 2024; Nadkarni et al., 2011). LLM's rapid advancement is due to increased data from IoT and accessible high-performance hardware (Johri et al., 2021). LLMs, like Gemini, are trained on vast text data, using transformer architectures to predict text sequences (Manning, 2022; Jana et al., 2024). Techniques like RLHF improve LLM output quality (Lambert et al., 2022). Gemini 1.5 Pro, used in this study, balances performance and accessibility (Gemini, n.d.-a, n.d.-b; Hassabis, 2024; Pichai, 2023, 2024; Reid et al., 2024).

**3.5.2 Large Language Model Applications in Data Extraction**

LLMs aid data science through AI agents and platforms like Google Cloud and Azure (Varshney, 2024; Guo et al., 2024; Ma et al., 2023). They excel in text summarization and coding assistance (Laskar et al., 2023; Chew et al., 2023). Prompt engineering, including agent simulations, few-shot prompting, and chain-of-thought reasoning, enhances LLM performance ("Prompting techniques", n.d.). Techniques like Word Embedding and Retrieval-Augmented Generation (RAG) are vital (CKIP Lab, n.d.; IBM, 2023; Lee, 2016;LlamaIndex, n.d.-a; *Retrieval augmented generation (RAG)*, n.d.; Meta AI, 2020.; Martineau, 2023). RAG tools like LlamaIndex and LangChain are used for efficient data retrieval (langchain-ai n.d.; Liu, 11 2022; Hugging face, n.d.; Gao et al., 2023; LlamaIndex, n.d.-b, n.d.-c). LLMs also assist in knowledge graph creation and integration (Pan et al., 2023; Microsoft, n.d.; Zhang et al., 2019; Knez & Žitnik, 2023; Neo4j, 2024).

**3.5.3 Large Language Model Applications in AI Speculative Design: Generating Data for SDG Roundtable Simulations**

AI's impact on SDGs is widely discussed (Vinuesa et al., 2020; Allen et al., 2024; Erdl, 2023; Fotopoulou, Mandilara, Zafeiropoulos, Laspidou, et al., 2022; Fotopoulou, Mandilara, Zafeiropoulos, & Papavassiliou, 2022). LLMs are used in speculative design scenarios, simulating human-AI interactions and generating research data (Lin & Long, 2023; Simeone et al., 2022; Mitrović, 2020; Rosenbaum et al., 2023; Aher et al., 2022). LLMs can simulate discussions, such as Socrates-style dialogues, to explore SDGs (Farnsworth, 2023). Roundtable discussions facilitate knowledge exchange (Rushmer et al., 2014). This study uses LLMs to create roundtable simulations, generating knowledge graphs to analyze SDGs and promote critical thinking.

# 4. Methodology

This section details the research methodology, which is divided into two parts. The first part focuses on processing text data from SDGs official indicator documents and TED Talks transcripts using LLMs and RAG. This process aims to create roundtable simulations and knowledge graphs to explore SDGs. The second part involves developing a visualization website to display the knowledge graphs generated from these simulations, facilitating further analysis and communication of the research findings (Figure 4-1).

**Figure 4-1**: System Data Architecture Diagram

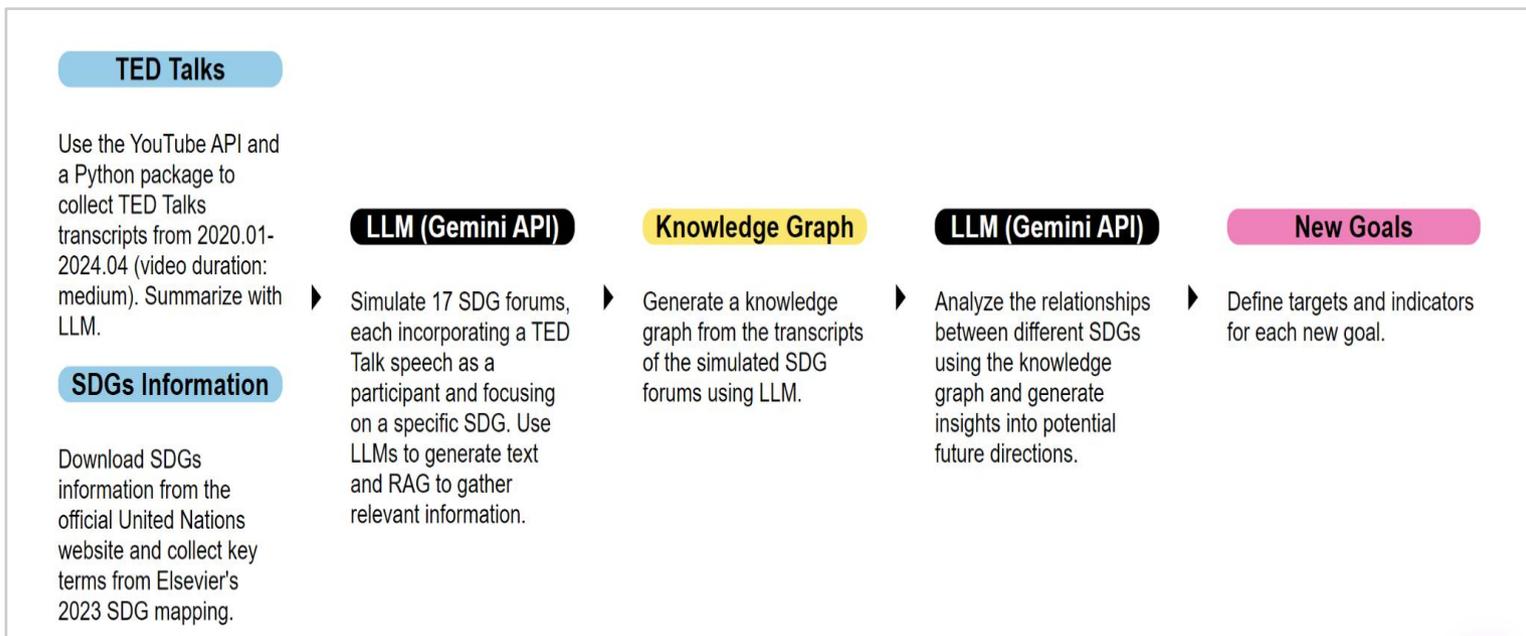

*Note:* The corresponding link to the "About" page of the website developed for this research is https://kg-web-4-0.vercel.app/about.html

## 4.1 Text Processing and System Flow Design
### 4.1.1 Research Data Text

The study utilizes two primary text data sources: SDGs official indicator documents (United Nations Statistics Division, n.d.-d) and TED Talks transcripts. The SDGs documents provide detailed information on each of the 17 goals, while TED Talks offer a diverse range of expert insights. TED Talks transcripts from videos published between January 2020 and April 2024 were selected for this study.

### 4.1.2 SDGs Basic Information and Keyword Integration

An SDGs database was created by processing the official SDGs indicator documents (Figure 4-2). The English version of the Excel file was used due to its compatibility with LLMs (Li et al., 2024). The data was cleaned and structured, with key information extracted and stored in a SQLite database. Keywords from Elsevier's SDGs academic research search engine (James, 2023) were added to enhance LLM processing.

**Figure 4-2**: SDGs Data Processing Workflow

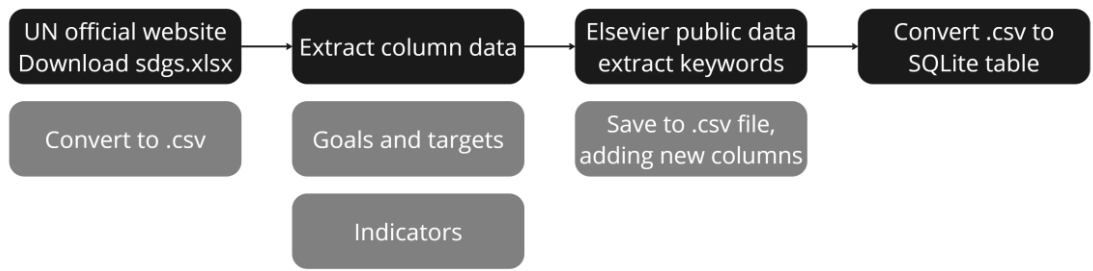

### 4.1.3 TED Talks Information Summarization via YouTube API and LLM

TED Talks video information was collected using the YouTube API (Youtube, n.d.) and transcripts were retrieved using the youtube-transcript-api Package (Depoix, n.d.). Following TED Talks' <18-minute guideline for effective knowledge delivery (Anderson, 2016) and to optimize Google YouTube API quota (excluding Shorts), we selected "medium" length videos (4-20 minutes per Google API documentation) to gather more comprehensive TED channel content (Google, n.d.-b; Anderson, 2016). The collected data was then summarized using LLM and RAG techniques (Figure 4-3). The LLM Gemini model (gemini-1.5-pro-latest) and the embedding model text-embedding-004 were employed. The summarization included key elements such as video description, core values, keywords, Q&A, and SDGs relevance (Table 4-1).

**Figure 4-3:** TED Talks Information Extraction, Identification, and Storage Workflow

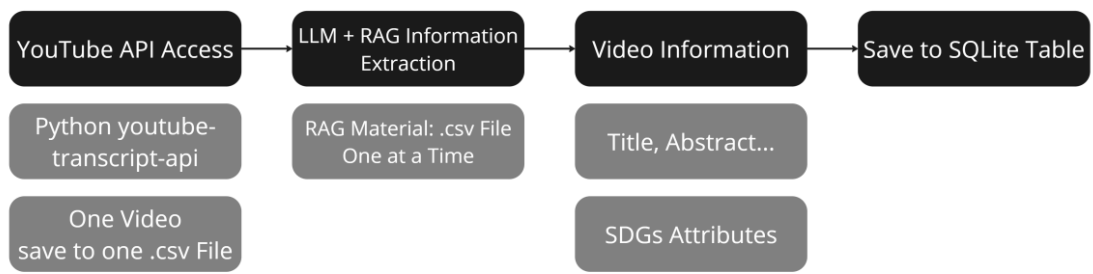

*Note:* Prompt: "You are a meticulous TED Talk summarization expert and social issue expert. Please carefully review the full text of tedtalk_data and accurately extract the following information: title, description, core_value, key_words. Additionally, generate 5 sets of Social Issue Q&A related to tedtalk_data, and analyze which SDG_types are highly relevant to tedtalk_data (no limit on quantity, but aiming for diversity and accuracy). Respond in English!"

**Table 4-1:** Item Content and Rationale for TED Talks Extracted by LLM.

| Item Name | Item Description |
| --- | --- |
| Video Summary (description) | Basic narrative summary compiled based on the video transcript. |
| Core Value of Video Content (core value) | Abstract and concrete knowledge core and knowledge value of the video. |
| Key Words (key words) | Representative words found within the video. |
| Five Sets of Social Issue Q&A (Q & A) | Considering the applicability and potential of the knowledge. |
| Corresponding SDGs Indicators (sdg type) | Revealing the SDGs types relevant to the knowledge. |

**4.1.4 Data Text Storage**

The processed TED Talks data was divided into two sets: a preliminary dataset (2023 data) and a formal dataset (2020.01-2024.04 data). The preliminary dataset, comprising 271 videos, was used for initial testing. The formal dataset, comprising 1127 videos, provided a more comprehensive dataset. Both sets were stored in SQLite databases (trans and trans2, respectively).

**4.1.5 Roundtable Simulation and Knowledge Graph Generation via LLM**

The study employed a two-phase approach. First, roundtable simulation transcripts were generated using LLM and RAG (Figures 4-4 and 4-5). Data was vectorized using RAG, and LLMs were prompted to generate roundtable transcripts simulating discussions focused on individual SDGs. These sessions featured anthropomorphized knowledge concepts engaging in quasi-Socratic dialogues to explore sustainability truths. Prompts designated a single SDG as the moderator, with relevant TED Talks

(identified via RAG based on shared SDG attributes) as participants. The LLM autonomously generated transcripts using the provided material, akin to human selection and was instructed to include a neutral facilitator to maintain focus and guide the discussion. The LLM Gemini model was used to generate transcripts, which were then used to create knowledge graphs (Figure 4-6). The knowledge graphs were stored in SQLite databases (forum and forum2).In the second phase, new SDGs were generated by overlaying the knowledge graphs (Figure 4-7). The LLM was prompted to identify relationships between SDGs and propose new goals, which were stored in SQLite databases (new goal and new goal2).

**Figures 4-4:** Single-Round Roundtable Simulation Diagram

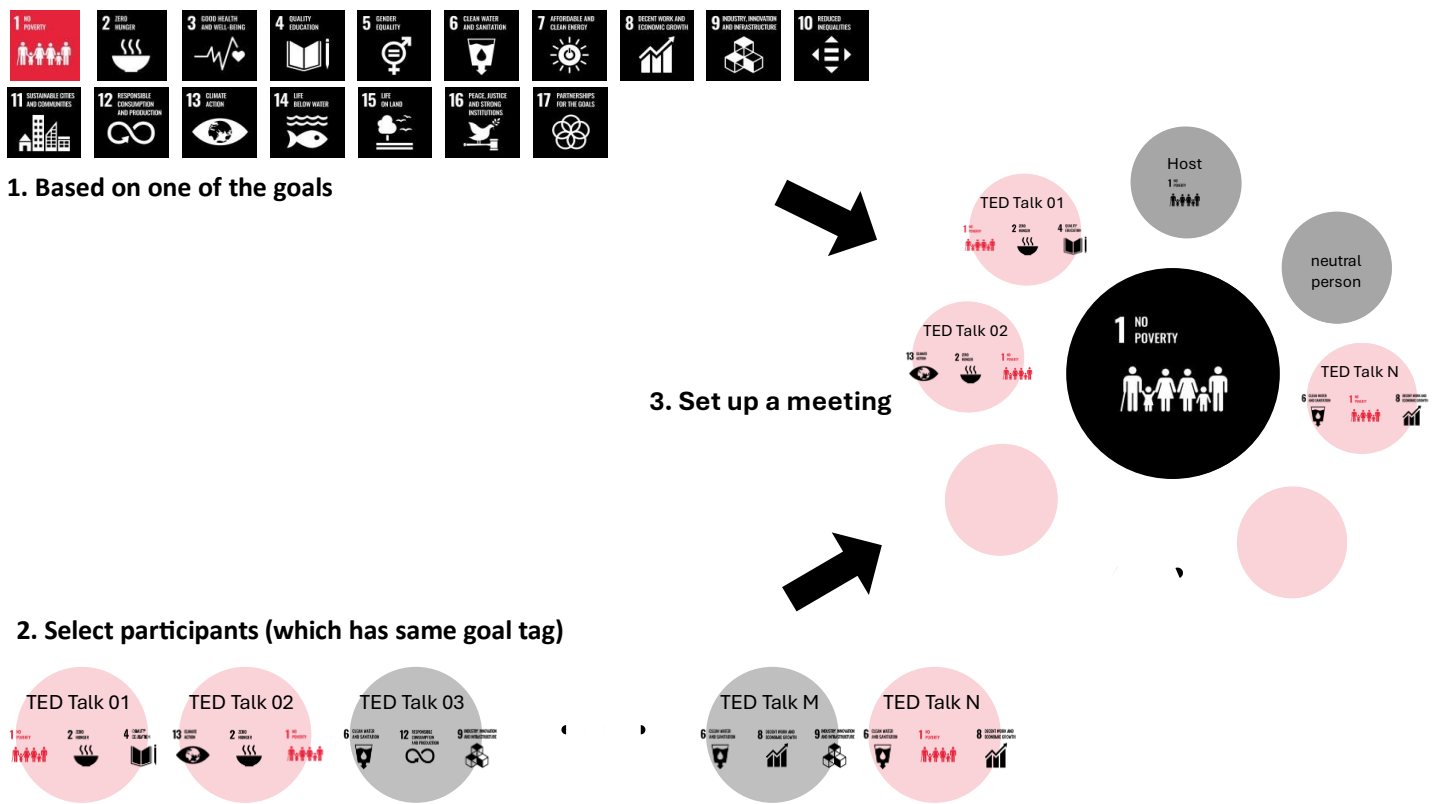

**Note:** A roundtable simulation meeting is formed by selecting corresponding participants from the database based on a single SDG goal.

**Figures 4-5:** Roundtable Simulation Transcript Generation Flowchart

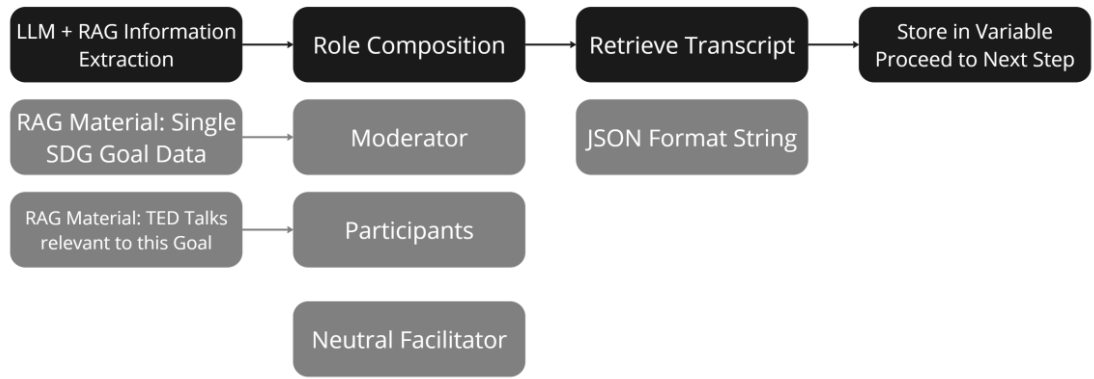

**Note:** Prompt: "You are a roundtable expert. Please simulate a roundtable meeting with anthropomorphized TED Talk knowledge and SDG concepts. The roundtable host is 【SDGs Goal "+self.type+"】, and the participants (members) are all the selected TED Talks (listed in the kg_box, where each TED Talk represents one participant, currently totaling "+n+" participants). Each participant will speak, with no limit on the total number of turns, and will fully express their stance, viewpoints, and the knowledge presented in their TED Talk, engaging in diverse, lively, and active multi-round questioning and debate. To ensure the meeting is concrete, valuable, and effective, a neutral facilitator will occasionally pose questions, including reminders to discuss more specific and feasible solutions, ranging from actions individuals can take to cooperation between governments and international bodies. All participants will then continue to actively discuss based on these questions until a clear and significant conclusion is reached (there will definitely be a conclusion). The discussion topic for this meeting is written in the bg_box (forum_background), which is to say, create a white paper based on the core values of 【SDGs Goal "+self.type+"】, focusing on implementation methods and innovative possibilities. The meeting will last for three hours. Please transcribe the complete meeting dialogue verbatim and respond in English (total word count: 8500 words)!"

**Figures 4-6:** Knowledge Graph Construction Workflow

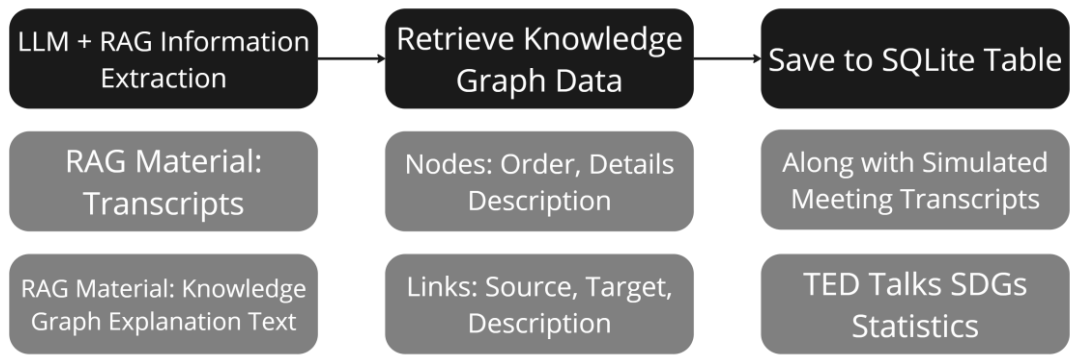

**Note:** Prompt: "You are a professional, reliable, precise, and highly enthusiastic knowledge graph construction expert. You excel at thoroughly reading roundtable meeting data and rigorously understanding the concept of ontology (described in kg_guide). Based on the transcript content (conversation_script) of the forum meeting and the ontology concept, meticulously identify and extract all nodes (id - node text content, order - the sequence in which it appears in the text, details - explanation of the node) and links (source - source content, target - target content, relation - relationship). The source and target of the links must definitively exist within the nodes and cannot be fabricated, while also aiming for richness of content. Present these nodes and links rigorously in the specified format. Respond in English! Ensure that the source and target are present within the nodes."

**Figures 4-7:** New SDGs Goal Generation Workflow

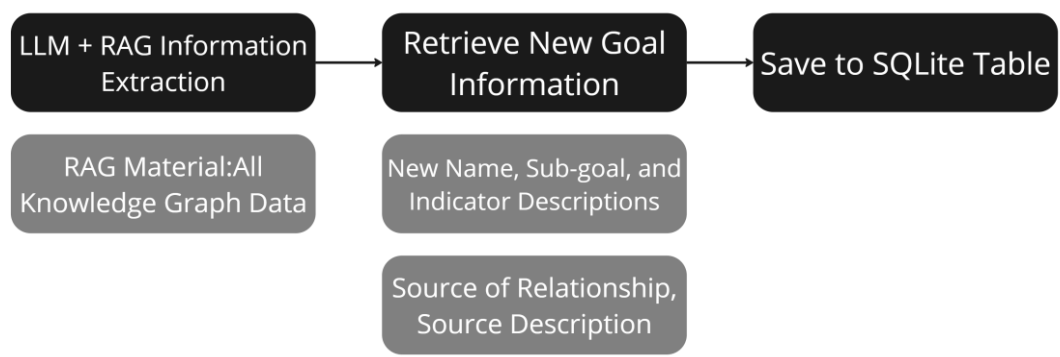

**Note:** Prompt: "You are a professional expert in knowledge graphs and thematic relationship analysis, skilled at analyzing the structure of knowledge graphs and the various relationships present within them. You can very carefully and accurately identify all the diverse relationships between SDGs goals, and simulate the generation of new SDGs goals based on the discovered relationships. Based on all the knowledge graph data in kg_data, very carefully identify all significant and important relationships between SDGs goals (no limit on the number of sets), list these relationships, and provide detailed explanations of their source (e.g., based on what knowledge graph content was it discovered). Finally, simulate the generation of new SDGs goals (starting from Goal 18 onwards), along with their targets and indicators in a format similar to this:【SDGs Goal 1. Content, abbreviated】 Answer in English, making the above content as diverse, rich, and accurate as possible, and place all results in 'results'."

## 4.2 Web Visualization

A website was developed to visualize the knowledge graphs and new SDGs generated in the first phase (Figure 4-8). The website includes pages for the SDGs knowledge graphs, new SDGs, and general information. The website uses HTML, CSS, and JavaScript, with D3.js for visualization and marked.js for Markdown rendering. The SDGs knowledge graph page displays the graphs in a spiral layout, with interactive elements for detailed information (Figure 4-9). The page also includes a control panel for adjusting the visualization. The new SDGs page presents the generated SDGs, with images and text descriptions (Figure 4-10). The website serves as a tool for visualizing the knowledge graphs and new SDGs.

**Figures 4-8:** Website Structure

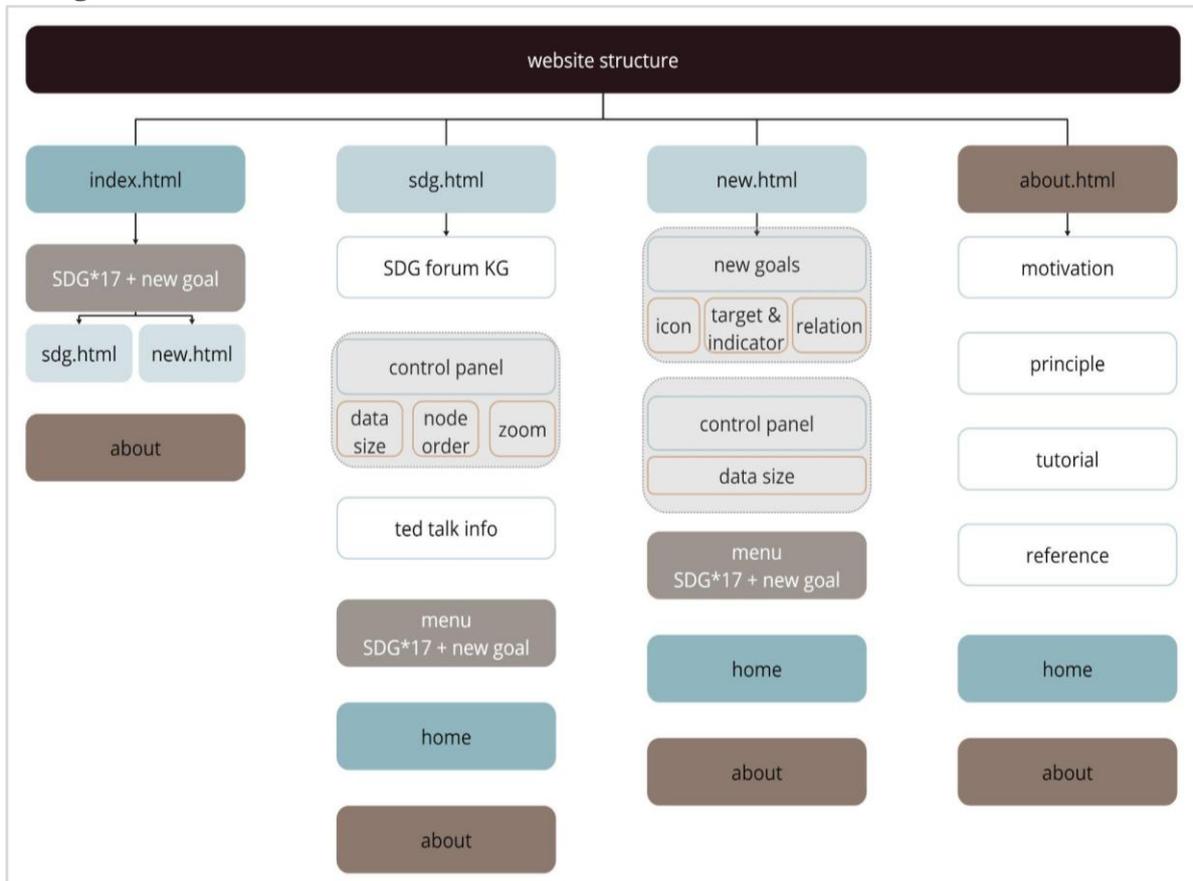

Note: Index.html is the homepage, sdg.html is the SDGs knowledge graph page, new.html is the new SDGs goals page, and about.html is the about page.

**Figures 4-9:** SDGs Knowledge Graph Page Screenshot

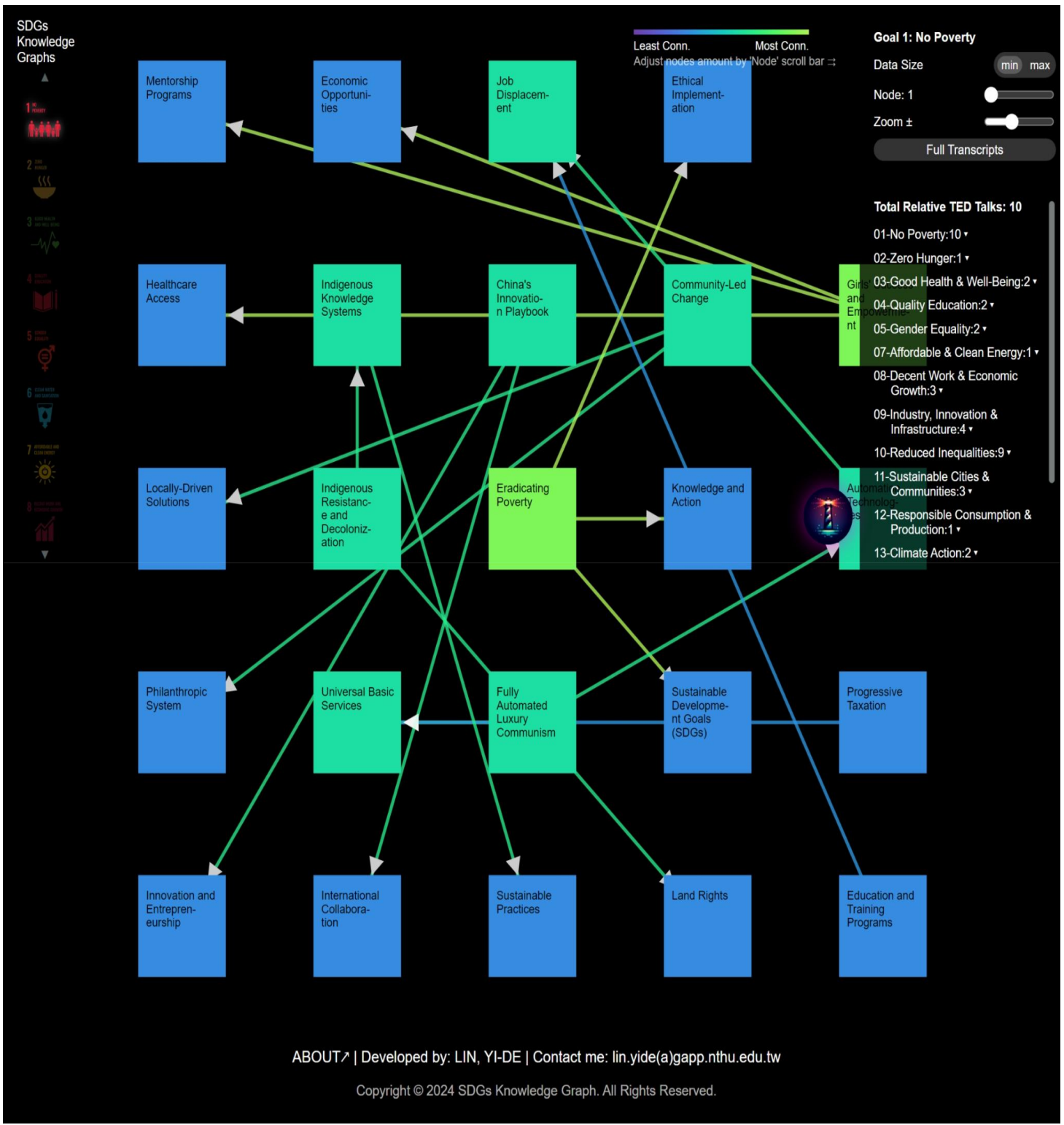

**Note:** The corresponding link to this page on the website developed for this research is https://kg-web-4-0.vercel.app/sdg.html. The control panel is on the right, and its functions are summarized in the table below.

**Figure 4-10:** New SDGs Goals Webpage Screenshot

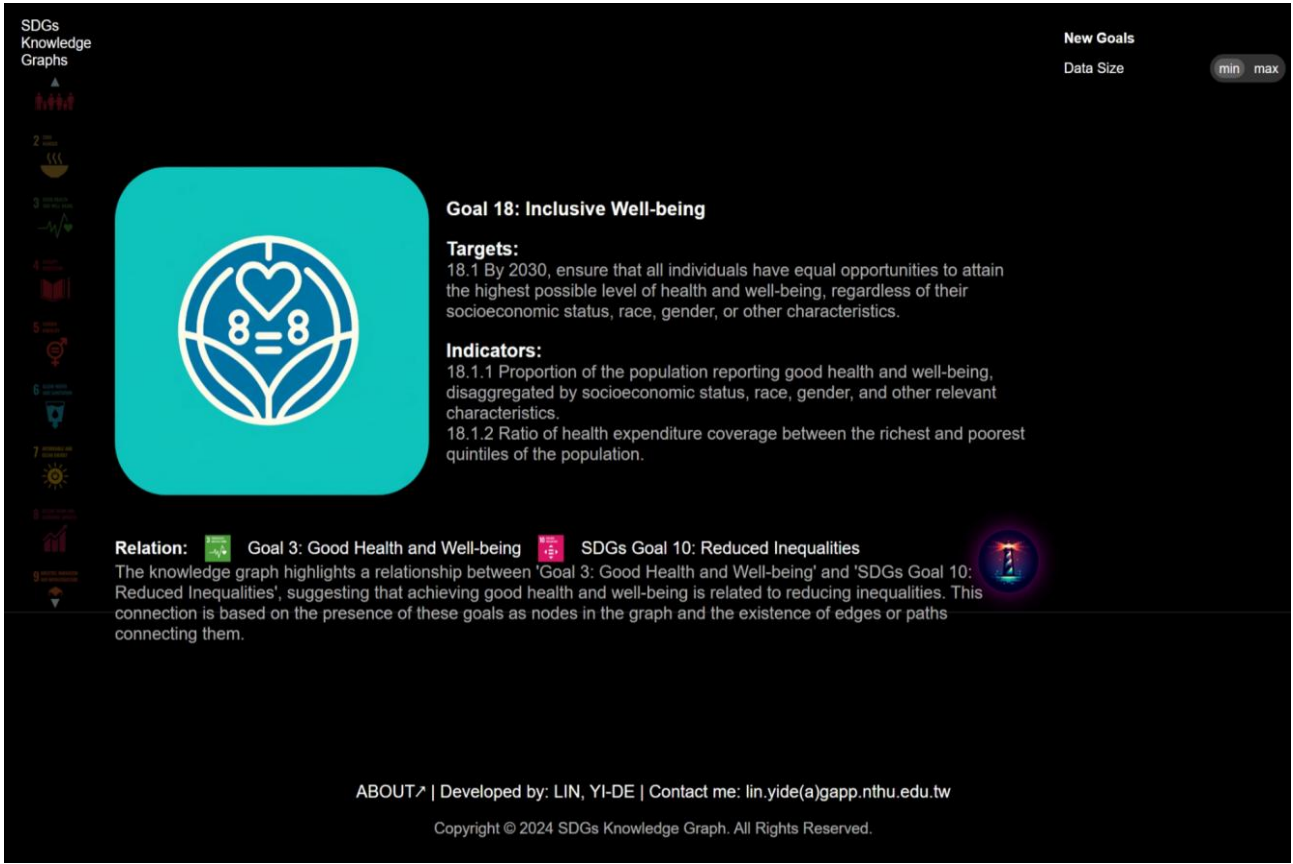

**Note:** The corresponding link to this page on the website developed for this research is:
https://kg-web-4-0.vercel.app/new.html

# 5. Research Results

## 5.1 Analyzing SDGs Composition in TED Talks to Explore Relationships Between SDG Goals

TED Talks gather diverse knowledge and ideas from global experts across multiple domains. Large Language Models (LLMs), trained on massive datasets, tend to reflect a microcosm of the world. This research adopts an AI-based inferential design approach, using LLMs to determine the SDG goal attributes of each TED Talk, examining SDG goal trends within our selected time frame, and further exploring relationships between SDG goals.

In the TED Talks database, the preliminary dataset was smaller in scale, covering 2023 with 269 usable entries containing 895 SDG goal attribute tags (averaging 3 SDG attributes per talk). The formal dataset was larger, spanning from January 2020 to April 2024, with 1,127 usable entries. After LLM analysis of each TED Talk's content for SDG goal attributes, we identified 3,730 SDG type tags across the dataset, also averaging 3 SDG attributes per talk.

Using SQLite queries, we filtered TED Talks with single SDG goal attributes, then calculated the number of talks corresponding to each of the 17 SDG goals. The data was visualized as heatmaps to identify which goals had stronger relationships from the perspective of TED Talks content, and compared with SDG goal relationships mentioned in previous literature (Pradhan et al., 2017).

### 5.1.1 Preliminary Dataset Heatmap Analysis

In the preliminary dataset heatmap (Figure 5-1), the central area (especially between Goals 7-13) showed higher values, suggesting these goals received more attention in our selected TED Talks. Edge areas (particularly between Goals 1-2, Goals 5-6, and Goals 14-15 corresponding to all goals) showed lower values, potentially indicating these goals received less emphasis in the talks.

The most frequent SDG attribute was Goal 10 (115 talks), followed by Goal 4 (102 talks), and Goal 16 (99 talks). Comparing with the United Nations' SDG framework indicator modifications over the years, Goal 16 (Peace, Justice & Strong Institutions) has undergone the most changes, while Goal 10 (Reduced Inequalities) tied for second most changes, further suggesting these two goals may receive higher attention in our TED Talks dataset.

**Figure 5-1:** Preliminary Dataset - Heatmap of SDGs Goal Attribute Tags for TED Talks Corresponding to Each SDG Goal

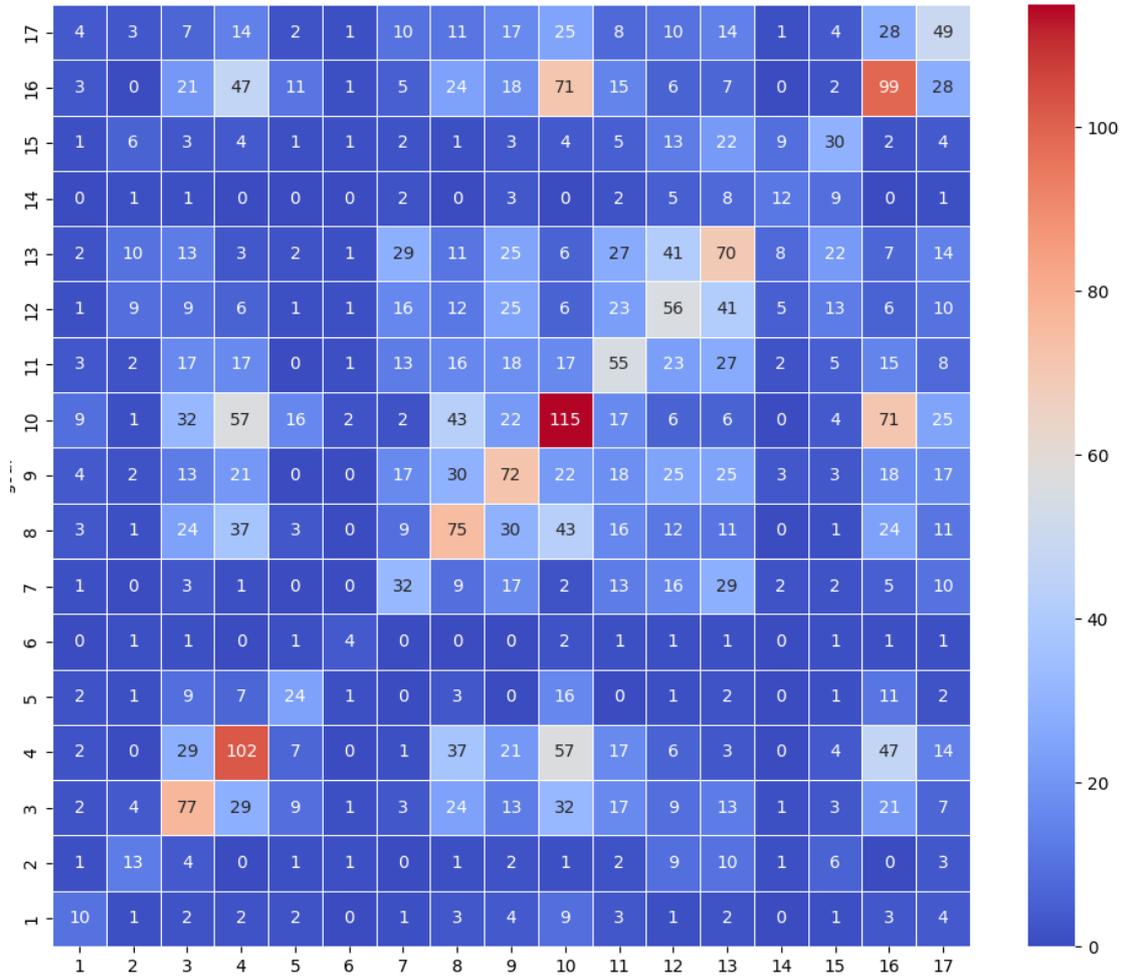

**Note:** The central region of the heatmap (especially between Goal 7 and Goal 13) shows higher values, while the peripheral regions (particularly between Goal 1 and Goal 2, Goal 5 and Goal 6, and Goal 14 and Goal 15 corresponding to all goals) show lower values.

Looking at TED Talks with two SDG goal attributes, the highest count was the combination of Goals 10 and 16 (71 talks), followed by Goals 4 and 10 (57 talks), and Goals 4 and 16 (47 talks). This data suggests that achieving certain goals may require considering potential relationships with other goals.

The edge areas of the heatmap showed lower values, with the minimum value being 0. This may be due to the limited size of the preliminary dataset (269 talks), resulting in more zero values, especially for talks with attributes of Goal 5 (Gender Equality), Goal 6 (Clean Water & Sanitation), and Goal 14 (Life Below Water), which appeared less likely to share attributes with other goals.

**5.1.2 Formal Dataset Heatmap Analysis**

The value distribution patterns were very similar between preliminary and formal datasets. In the formal dataset heatmap (Figure 5-2), the central area (especially between Goals 7-13) showed higher values, while edge areas (particularly between Goals 1-2 corresponding to all goals) showed lower values.

**Figure 5-2:** Formal Dataset- Heatmap of SDGs Goal Attribute Tags for TED Talks Corresponding to Each SDG Goal

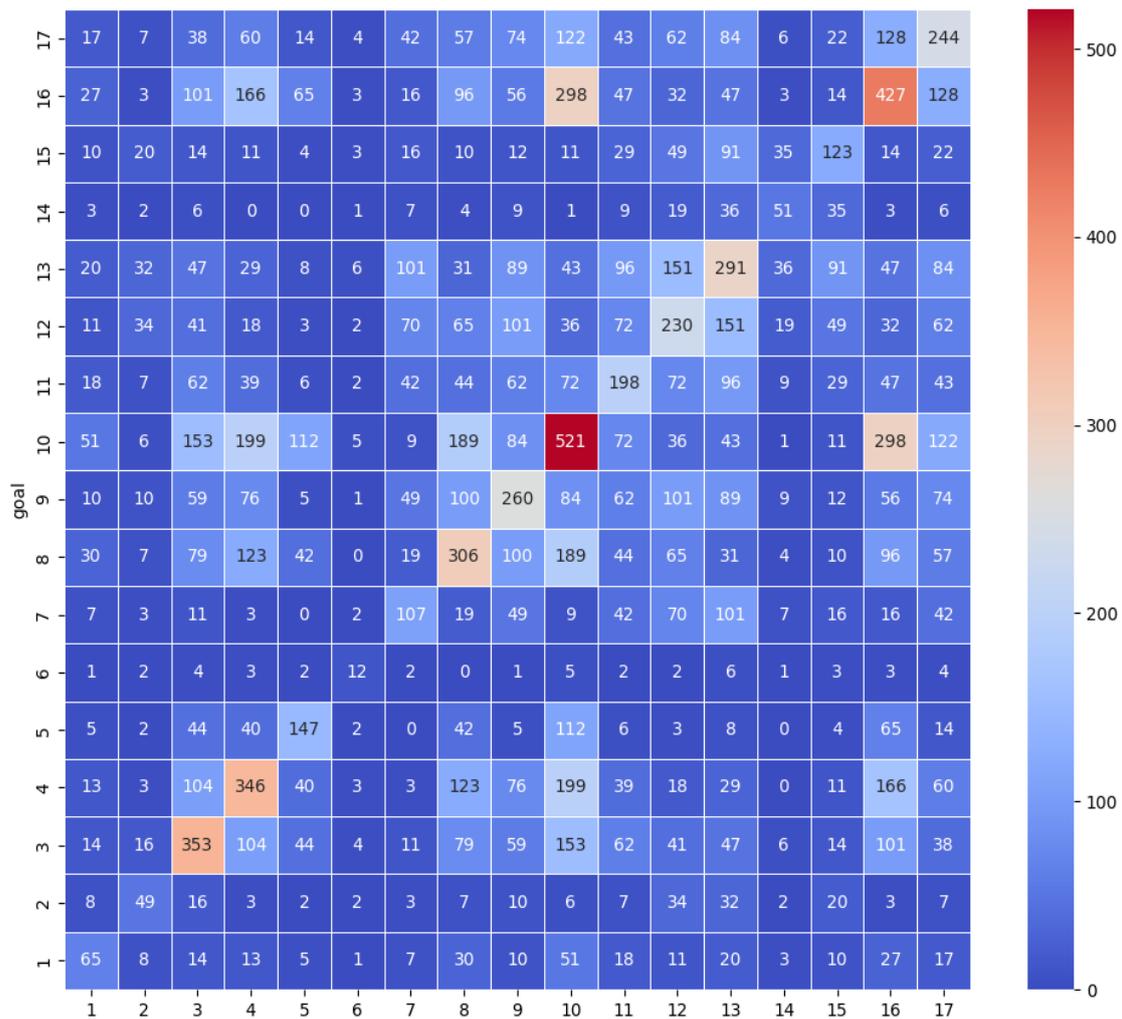

**Note:** The central region of the heatmap (especially between Goal 7 and Goal 13) shows higher numerical values, while the peripheral regions of the heatmap (particularly between Goal 1 and Goal 2 corresponding to all goals) show lower numerical values.

The highest value was for Goal 10 (521 talks), followed by Goal 16 (427 talks), and Goal 3 (353 talks). In other words, these three SDGs dominated the formal dataset, characterizing the attribute tendencies of our TED Talks data. Additionally, Goals 10 and 16 ranked in the top three in both the preliminary and formal datasets, potentially validating, as mentioned in the preliminary analysis, that these two goals receive higher attention globally when compared with the frequency of UN indicator modifications to the SDG framework.

For TED Talks with two SDG attributes, the highest count was the combination of Goals 10 and 16 (298 talks), followed by Goals 4 and 10 (199 talks), and Goals 8 and 10 (189 talks). The second highest combination, Goals 4 (Quality Education) and 10 (Reduced Inequalities), showed a positive correlation in previous research (Pradhan et al., 2017), suggesting a relationship worth further exploration. The consistency of the top two combinations between preliminary and formal datasets further validates that our selected TED Talks have representative qualities, with the preliminary dataset often representing the formal dataset, while the formal dataset more appropriately represents global knowledge contexts and phenomena.

The lowest value areas in the heatmap (Figure 5-2) had values of 0, including intersections between Goals 4 and 14, Goal 5 with Goals 7 and 14, and Goals 6 and 8. This suggests our TED Talks data less frequently contained these attribute combinations, or these goal combinations have lower relationships. For example, similar patterns in the preliminary dataset showed Goals 5 and 14 had less likely associations with other goals, with more frequent zero values when combined with other goals.

From another perspective: 346 TED Talks had Goal 4 attributes, 147 had Goal 5 attributes, and 51 had Goal 14 attributes. From our TED Talks data perspective, with relatively fewer Goal 14 (Life Below Water) attribute tags, Goals 4 (Quality Education) and 5 (Gender Equality) showed little relationship with Goal 14. Similarly, only 12 talks had Goal 6 attributes (the least frequent), while 306 had Goal 8 attributes. This could indicate not only a low relationship between Goals 6 (Clean Water & Sanitation) and 8 (Decent Work & Economic Growth), but also that recent TED Talks less frequently discuss Goal 6 topics. While Goals 7 (107 talks) and 5 (147 talks) had similar frequencies, they never appeared together in the same talk (0 intersections), suggesting minimal relationship between Goals 5 (Gender Equality) and 7 (Affordable & Clean Energy).

Further visualizing SDG goal attributes as nodes in a network graph, with connections between attributes appearing in the same TED Talk (Figure 5-3), reveals closer relationships between certain SDG goals. For instance, Goal 10 (Reduced Inequalities) appears to require integration with Goals 16 (Peace, Justice & Strong Institutions), 4 (Quality Education), and 8 (Decent Work & Economic Growth). Research shows certain groups in disadvantaged positions more easily encounter economic difficulties and unfair situations, requiring fair and just organizational units to provide decent work, economic support, and comprehensive educational measures to help them overcome challenges (Mabuza, 2020).

**Figure 5-3:** Formal Data Text - Network Graph of SDGs Goal Attributes for TED Talks

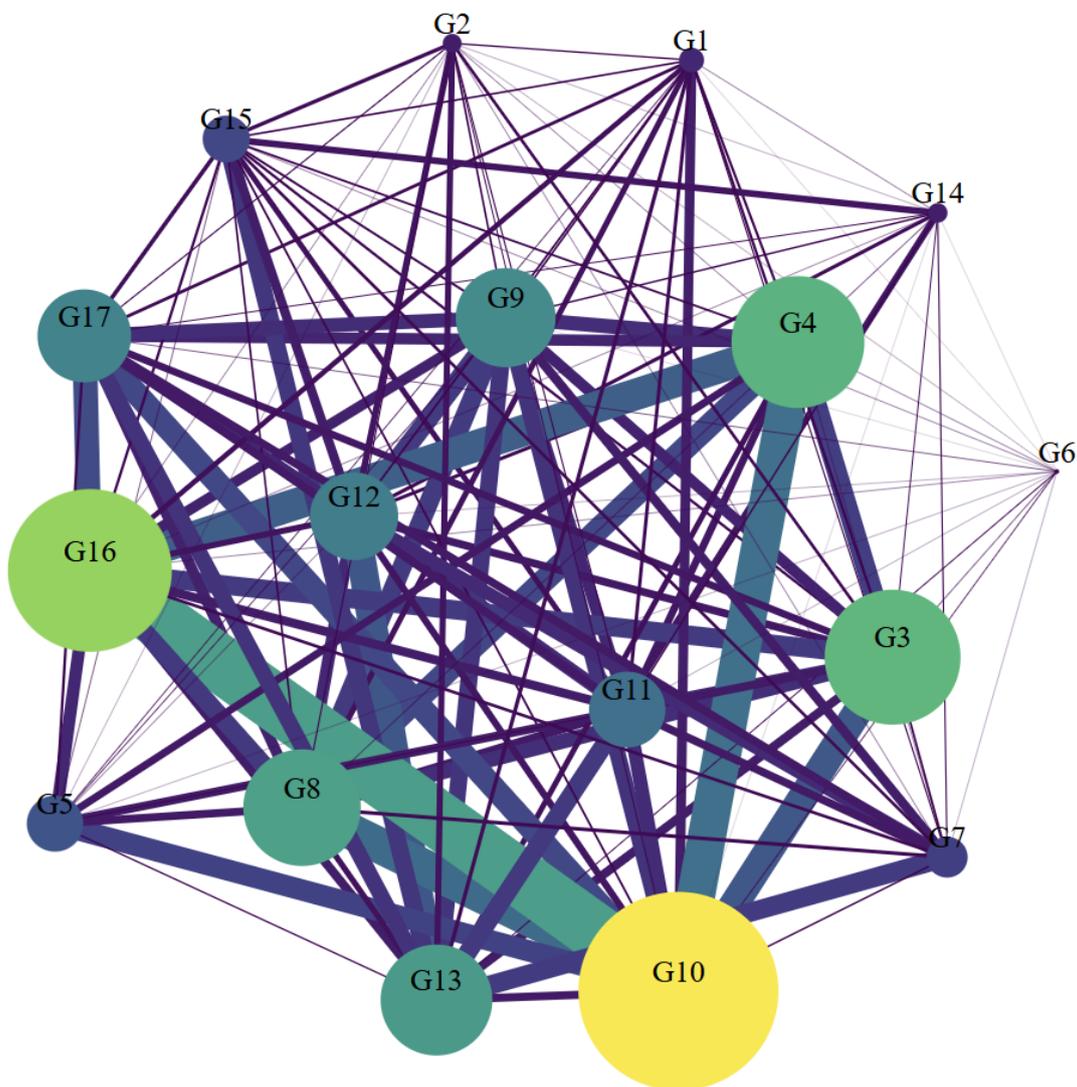

**Note:** G is an abbreviation for Goal; for example, G10 represents Goal 10. This graph displays data from the website developed for this research.

If our TED Talks data adequately represents the current state of global knowledge and ideas, then the absence of certain goal combinations in our data indicates a global lack of knowledge and ideas in these areas. This can also serve as an extension to previous literature on relationship scales between SDG goals from an integrated policy perspective (Nilsson et al., 2016), highlighting that integrating certain goal combinations into policy guidelines has few precedents, presenting both new challenges and opportunities. The heatmap not only shows the relational status of specific SDG goal combinations but also illustrates trends in our TED Talks dataset, or indirectly shows how SDG goal attributes and prompt content influence system construction in AI inferential design.

## 5.2. Visual Synthesis of Knowledge Graphs for Each Goal

After visualizing knowledge graph structural data obtained through an AI inferential design approach, we explored the implicit meanings and key elements of each SDG goal. We observed structural information through text and image composition trends, analyzing three aspects: first, comparing the initial nodes (earliest concepts) with nodes having the most connections in each knowledge graph to identify the truly critical elements; second, examining node color distribution patterns to observe the visual manifestation of simulated roundtable discussions; third, synthesizing the directional tendencies of connecting arrows to understand relationship trends between core nodes and other nodes.

### 5.2.1 Comparison Between Initial Nodes and Most Connected Nodes

In most knowledge graphs, the initial node and the node with the most connections were identical. The initial node typically represented the roundtable discussion topic (a specific SDG goal), with other concept nodes radiating outward from this starting point. However, in some knowledge graphs, the initial node differed from the most connected node, suggesting that certain concepts within that SDG-themed roundtable discussion warranted deeper investigation. To understand the significance of this difference, we analyzed these specific knowledge graphs in detail.

In the formal dataset, Goal 4 (Quality Education) and Goal 6 (Clean Water & Sanitation) had "White Paper" as their most-connected node, reflecting that these goals particularly require clear documentation; Goal 5 (Gender Equality) had "Implementation Strategies" as the most-connected node, indicating a need for more explicit execution strategies; Goal 7 (Affordable and Clean Energy) had "Progress Measurement" as an additional most-connected node, suggesting that effective monitoring methods are needed for progress; Goal 9 (Industry, Innovation & Infrastructure) had "Holistic Approach" and "Ethical Imperative" as additional most-connected nodes, highlighting the need for comprehensive, highly collaborative approaches that consider ethical implications; Goal 10 (Reduced Inequalities) had "Individual Action" as its most-connected node, emphasizing the importance of personal actions including confronting bias-related challenges and supporting various organizations and equity policy advocacy; Goal 16 (Peace, Justice & Strong Institutions) had "Disinformation" as its most-connected node, highlighting the need to address the impacts of misinformation.

In the preliminary dataset, Goal 1 (No Poverty) had "Eradicating Poverty" and "Girls' Success and Empowerment" as most-connected nodes, aligning with research showing disadvantaged groups often face multiple rights violations (Mabuza, 2020) and suggesting that addressing vulnerable group issues may help secure fundamental rights; Goal 2 (Zero Hunger) had "Sustainable Rice Farming" as an additional most-connected node, presenting an interesting perspective that sustainable rice cultivation methods could address hunger issues; Goal 6 (Clean Water & Sanitation) had "Sustainable Living" as its most-connected node, indicating that achieving this goal requires lifestyle changes including water conservation and responsible consumption; Goal 8 (Decent Work & Economic Growth) included "Data Ownership," "Boredom," and "AGI" as additional most-connected nodes, highlighting the importance of personal data control rights and psychological states in economic development; Goal 17 (Partnerships for the Goals) included "Online Communities" as an additional most-connected node, suggesting that online communities can facilitate goal achievement.

Synthesizing these findings, the most-connected nodes from the formal dataset generally related to broader implementation guidelines, strategies, and organizational roles, while those from the preliminary dataset often represented more novel solutions and conceptual terms.

**5.2.2 Node Color Distribution in Knowledge Graphs**

Node color is determined by the spectrum color corresponding to the number of connections, with the most connected nodes appearing lightest and the least connected nodes appearing darkest. Observing the color distribution in knowledge graphs (Figure 5-4), in most cases, lighter-colored nodes clustered in the inner layers (earlier nodes) with more connections, while darker-colored nodes appeared in the outer layers with fewer connections. However, in some knowledge graphs, nodes with more connections appeared in the outer layers, suggesting that certain topics required later-stage discussion to reveal concepts that interconnect with multiple nodes. This pattern, revealed through our AI inferential design approach to simulated roundtable discussions, demonstrates that these SDG goals possess attributes requiring multi-layered discussion, application, and analysis in actual society.

**Figure 5-4:** Overview of Knowledge Graphs

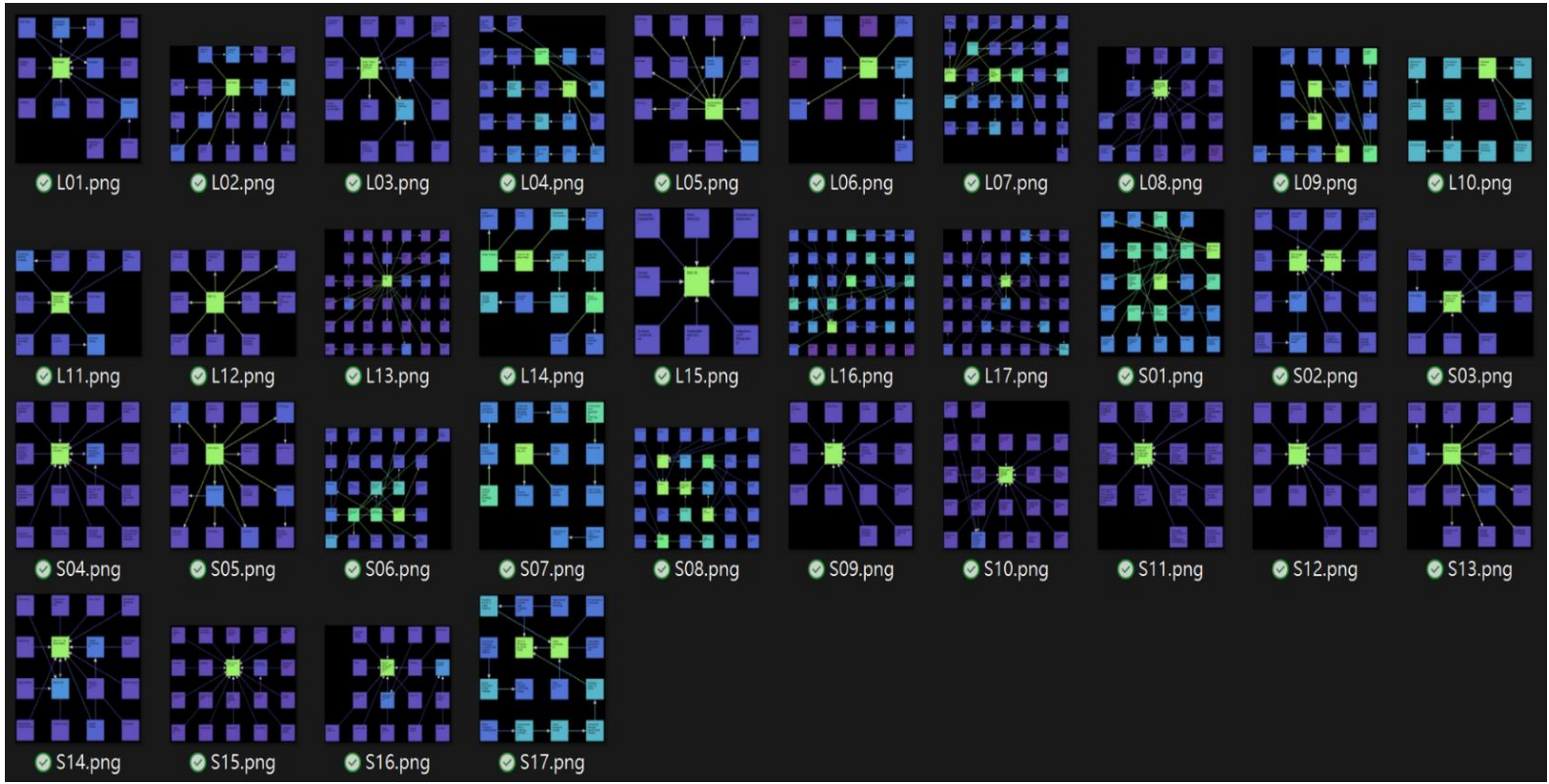

**Note:** The image code L represents formal data text (1127) derived from a larger number of TED Talks videos, and the image code S represents pilot data text (269) derived from a smaller number of TED Talks videos. The numerical codes represent the respective SDGs goal codes; for example, L01 refers to the knowledge graph of Goal 1 from the formal data text.

Examining knowledge graphs where the most-connected nodes appeared in the outer layers—Goal 7 (Affordable & Clean Energy), Goal 9 (Industry, Innovation & Infrastructure), Goal 10 (Reduced Inequalities) from the formal dataset, and Goal 1 (No Poverty) from the preliminary dataset—we found these also exhibited differences between initial nodes and most-connected nodes. This suggests that multi-layered discussions more easily lead to situations where initial nodes differ from most-connected nodes, implying that deeper critical points emerge through more extensive discussion.

Regarding node color distribution, the knowledge graph with the most color varieties was Goal 17 (Partnerships for the Goals) from the formal dataset (labeled L17 in Figure 5-4), displaying 7 colors across 41 nodes and 37 connections. This indicates a lively, detailed simulated roundtable discussion that produced interconnected concepts worthy of further exploration. The knowledge graph with the lightest overall node colors was Goal 10 (Reduced Inequalities) from the formal dataset, with only 3 node colors across 12 nodes and 6 connections, with nodes forming several small clusters through connections, suggesting insufficient discussion cohesion. The knowledge graphs with the fewest node colors were several from the preliminary dataset (S9, S11, S12, S15), displaying only the lightest and darkest colors. This possibly indicates insufficient data for these roundtable simulations, though when considering their most-connected nodes, they tended to generate distinctive and concrete solutions, resulting in all nodes connecting to the initial theme node but with limited interconnections.

**5.2.3 Directional Trends of Connection Arrows in Knowledge Graphs**

The directional trends of connection arrows in knowledge graphs generally fell into two categories: inward pointing, where outer (later) nodes pointing to inner (earlier) nodes exceeded inner nodes pointing to outer nodes; and outward pointing, where outer nodes pointing to inner nodes were fewer than inner nodes pointing to outer nodes. Considering the semantic directionality of knowledge graph connections from subject to object or more concrete descriptions (Ehrlinger & Wöß, 2016; IBM Technology, 2022; Xiao, 2021), when arrows pointed inward, outer (later) nodes were generally more abstract than inner (earlier) nodes, indicating more abstract, divergent thinking in the simulated roundtable discussions; when arrows pointed outward, outer (later) nodes were more concrete than inner (earlier) nodes, indicating more concrete, focused discussions.

In the formal dataset, seven knowledge graphs exhibited characteristics of having different initial and most-connected nodes with outward-pointing arrows. Four of these—Goal 5 (Gender Equality), Goal 6 (Clean Water & Sanitation), Goal 7 (Affordable & Clean Energy), and Goal 9 (Industry, Innovation & Infrastructure)—were rated as "Moderately Improving" in SDG progress trends by scholars (Sachs et al., 2023). For Goals 7 and 9, the most-connected nodes appeared in the outermost layers of their knowledge graphs, suggesting that roundtable discussions on these topics generated critical points distinct from the initial themes through multi-layered, divergent discussions. This reflects that these goals' multi-layered ideation potential in the real world contributes to better implementation status, revealing how the difficulty of framework-to-solution concretization affects goal achievement (Kim, 2023).

Knowledge graphs from the formal dataset more frequently displayed outward-pointing arrows and differences between initial and most connected nodes compared to those from the preliminary dataset. The preliminary dataset's knowledge graphs more often exhibited inward-pointing arrows and node color distributions limited to the lightest and darkest colors (Appendix 2). This phenomenon suggests that when conducting SDG roundtable simulations from an AI inferential design perspective, larger data volumes (as in the formal dataset) more easily generate concrete, focused ideation, while smaller data volumes (as in the preliminary dataset) more readily produce abstract, divergent thinking. This implies that in actual society, effective improvement of the SDG framework requires sufficient data collection to support optimization proposal ideation and precisely identify core issues.

### 5.3 New SDGs Goals Content and Structure

After analyzing knowledge graphs from preliminary and formal text data through LLM, new goals were generated based on identified relationships. The preliminary text data yielded 1 new goal, while the formal text data produced 5 new goals.

### 5.3.1 Preliminary Text Data: One New Goal

**1. Inclusive Well-being**: Combining Goal 3 and Goal 10, this aims to ensure everyone has equal access to the highest standards of health and well-being regardless of socioeconomic status, race, gender, or other characteristics. Indicators track the proportion of the population reporting good health and well-being disaggregated by various characteristics, and the ratio of health expenditure coverage between the richest and poorest population quintiles.

This new goal reflects the close relationship between Goal 3 (Good Health & Well-Being) and Goal 10 (Reduced Inequalities) frequently appearing in knowledge graphs, confirming the positive correlation highlighted in research (Pradhan et al., 2017) and supporting findings that inequality exacerbates risks of well-being deprivation (Mabuza, 2020).

### 5.3.2 Formal Text Data: Five New Goals

**1. Poverty Reduction through Technological Advancement**: Combining Goal 1 and Goal 9, this aims to utilize and provide affordable innovative technologies to accelerate poverty elimination, bridge technological gaps, and promote economic development. Indicators include the proportion of digital devices owned by impoverished populations and the percentage of people living in poverty.

**2. Climate Resilience for Vulnerable Communities**: Combining Goal 1 and Goal 13, this focuses on strengthening the resilience of impoverished populations facing extreme climate disasters. Indicators track the number of plans enhancing climate change resilience for impoverished populations and the proportion of impoverished households receiving climate disaster warning information.

**3. Inclusive and Equitable Development**: Combining Goal 5 and Goal 10, this aims to eliminate various forms of inequality and discrimination while promoting equal opportunities for diverse groups. Indicators include pay differences across genders in various sectors and the proportion of leadership positions held by women and marginalized groups.

**4. Global Collaboration for Water Security**: Combining Goal 6 and Goal 17, this aims to establish a global water security alliance to promote cooperation between countries, organizations, and social groups in addressing water resource challenges. Indicators track the number of international agreements on water security and funding allocated to collaborative water resource management projects.

**5. Inclusive Economic Empowerment**: Combining Goal 8 and Goal 1, this promotes inclusive and sustainable economic growth with appropriate work opportunities for everyone, especially impoverished populations. Indicators include employment rates and average personal income of impoverished populations.

Among the new goals from formal text data, Goal 1 (No Poverty) was referenced three times, while others were referenced once each. This emphasis on poverty issues wasn't directly observed in TED Talks SDGs attribute statistics or roundtable simulation knowledge graphs but aligns with research showing Goal 1 correlates positively with many other SDGs goals (Pradhan et al., 2017).

Although the total number of goals differs between preliminary and formal text data, each goal contains one sub-goal and two indicators regardless of data source, despite this not being specified in the prompt. Both data types referenced Goal 10 (Reduced Inequalities), possibly reflecting its prevalence in TED Talks data or AI's recognition of its societal importance. All referenced goals are currently facing "Major challenges" or "Significant challenges" according to research (Sachs et al., 2023), indicating they require greater collaborative relationships or new framework directions to overcome difficulties.

# 6. Conclusion

This research constructed an SDGs knowledge graph system from an AI speculative design perspective, aiming to create an imaginative space for discussing future sustainable development and answering three research questions:

## 6.1. Question 1

We examined the SDGs attribute composition trends in TED Talks videos to understand relationships between various SDGs goals. Analysis of the SDGs target attribute markup heat map revealed that: (1) Goal 10 was the most prevalent attribute across all TED Talks data, indicating greater emphasis on this discussion in both TED Talks and potentially the real world. Among content with Goal 10 attributes, most simultaneously featured Goal 16 attributes, demonstrating a connection between Goal 10 (Reduced Inequalities) and Goal 16 (Peace Justice & Strong Institutions) that requires mutual consideration in practical applications. (2) Goal 6 was the least represented attribute across all TED Talks data, and no content with Goal 6 attributes simultaneously featured Goal 8 attributes, suggesting minimal connection between these goals or indicating a knowledge and idea gap in TED Talks and potentially worldwide discourse.

## 6.2. Question 2

We explored the structure, content, and visualization outcomes of SDGs roundtable simulation knowledge graphs from a speculative design perspective to identify key SDGs elements. By comparing initial nodes with most-connected nodes, analyzing color distribution and types after visualization, and examining directional tendencies of node connections, the research found that when initial nodes differed from most-connected nodes, the latter held significance worthy of deeper investigation. This phenomenon appeared in Goals 4, 5, 6, 7, 9, 10, and 16 in formal text data. Most-connected nodes typically discussed broader implementation guidelines and methods, while the most-connected nodes in preliminary text data for Goals 1, 2, 6, 8, and 17 often discussed novel concepts and terminology. Among knowledge graphs where initial and most-connected nodes differed, the most-connected nodes in formal text data for Goals 7, 9, 10 and preliminary text data for Goal 1 appeared in the outer layers, indicating that extended discussion tends to produce potential key nodes distinct from initial nodes. Additionally, arrows in formal text data knowledge graphs tended to point outward, while those in preliminary text data pointed inward, suggesting that in AI speculative design, larger data volumes mapped to appropriate goals (e.g., Goals 7 and 9 in formal text data) facilitate concrete ideation and better implementation outcomes.

## 6.3. Question 3

We attempted to generate new SDG goals by overlaying knowledge graphs and exploring potential goal relationships. The research found that all preliminary text data knowledge graphs generated one new goal, "Inclusive Well-being," emphasizing equal access to high-quality health and well-being for all, referencing Goals 3 and 10 and validating their correlation as important reference points for new SDGs frameworks. All formal text data knowledge graphs generated five new goals: "Poverty Reduction through Technological Advancement" (bridging technological gaps between rich and poor), "Climate Resilience for Vulnerable Communities" (emphasizing equal resilience to climate change), "Inclusive and Equitable Development" (emphasizing diverse equality), "Global Collaboration for Water Security" (emphasizing cooperative approaches to water resource challenges), and "Inclusive Economic Empowerment" (emphasizing work and economic growth opportunities to eliminate poverty). This demonstrates the system's ability to discover new goal directions through multi-goal relationship mining, validating that Goal 1, the most frequently referenced SDGs goal, correlates with many other SDGs goals. Comparison with the system's prompt reveals that AI speculative design execution is influenced by prompt content but maintains a degree of uncertainty. Both preliminary and formal text data referenced Goal 10, reaffirming its real-world importance.

In conclusion, this research used TED Talks data and LLM roundtable simulations to generate knowledge graph data, visualizing and deeply exploring SDGs goal relationships and internal structures. It discovered deep-level connections between different goals, such as the key role of vulnerable groups in realizing SDGs goals and the interconnected nature of goals requiring holistic consideration for optimal results. By generating new goals, it provided a fresh perspective for discussing and understanding sustainable development issues, serving as an AI speculative design experiment using methods previously unapplied in SDGs development research. This research's imaginative space explores whether the framework constraining human development needs reconsideration.

# 7. Discussion

## 7.1 Differences Between Two TED Talks Data Sets

The formal text data represents TED Talks video transcripts from 2020.01-2024.04 processed by LLM, comprising 1127 usable entries, while the preliminary text data represents similarly processed TED Talks from 2023, comprising 269 usable entries. The preliminary data was used for the research's predictive test procedures, and the formal data for the official procedures. This allowed us to explore whether different data scales affect the system constructed from an AI speculative design perspective.

### 7.1.1 Supplementary Comparison of Knowledge Graph Nodes and Links

The formal text data knowledge graphs contained 343 total nodes and 303 links, while preliminary text data knowledge graphs contained 301 nodes and 276 links. The formal text data averaged 20.176 nodes and 17.824 links per SDGs goal knowledge graph, while preliminary text data averaged 17.706 nodes and 16.235 links.

Analysis of knowledge graph nodes showed that Levene's test was significant, indicating heterogeneity of variance. Using an independent samples t-test assuming unequal variances, no significant difference was found ($p = 0.415$) between formal and preliminary text data knowledge graph node counts.

Similarly, analysis of knowledge graph links showed significant Levene's test results, indicating heterogeneity of variance. Using an independent samples t-test assuming unequal variances, no significant difference was found ($p = 0.570$) between formal and preliminary text data knowledge graph link counts.

### 7.1.2 Integrated Comparison and Significance

1. From the SDGs attribute analysis heat maps, the distribution trends and identified SDGs relationships were very similar between preliminary and formal text data. This suggests the preliminary text data had sufficient representativeness for SDGs attribute analysis, though formal text data filled in some numerical gaps.

2. For knowledge graphs: (1) Statistical methods verified no significant differences in node and link totals between preliminary and formal text data. (2) However, formal text data more frequently produced situations where the most-connected nodes representing key content differed from initial nodes, and showed more complex arrow direction structures pointing outward. This indicates different complexity levels and thematic development directions between the two data sets.

3. For new SDGs goal content: (1) Without specific prompt instructions, both produced goals with identical numbers of sub-goals and indicators. (2) However, formal text data produced more new goals and revealed more goal relationships compared to preliminary text data, with different content emphases.

In other words, preliminary and formal text data differed in content but were similar numerically. Content differences may result from LLM or prompt influences, reflecting AI's uncertain variability, plus differences in textual detail due to different data volumes. Through multiple rounds of LLM understanding and summarization, latent textual features in the data were gradually amplified, resulting in noticeable differences in generated content. However, numerical aspects including knowledge graph node and link counts and new SDGs goal structure showed similarities likely related to data format, possibly influenced by LLM architecture, constraints, or prompt word limits.

During the research process, we found that excessive data volume for RAG processing failed to enhance weighting and potentially caused loss of key information, an issue noted by some LLM application developers (RAG is failing when the number of documents increase, 2024; Zafar, 2024). This explains why only TED Talks data itself was used for LLM SDGs goal attribute determination, without additional SDGs framework data. Therefore, the fundamental principle of LLM's interpretation of TED Talk SDGs attributes stems from the LLM's inherent tendencies or model randomness, requiring further repeated experiments and sample expansion for validation.

### 7.2 Research Limitations

TED Talks videos contain elements including images, sound, and text. While textual information can be obtained from subtitles, presenters' slides, tone, and body language cannot be directly extracted from subtitles. This limitation could be addressed in future research through multimodal artificial intelligence.

Additionally, some TED videos on YouTube are member-exclusive, preventing direct scraping of complete transcripts to protect consumer rights and copyright. This limited information to titles and description content. Future research might consider direct collaboration with TED after developing more complete system architecture and research content.

Finally, the data used in LLM model construction, safety regulations, response instability, and prompt formulation all affected LLM-generated data. Some TED Talks data potentially involved safety-related issues, hindering summary extraction. Since

LLM training derives from human information patterns, the model may inherit undiscovered human blind spots.

### 7.3 Future Development

While many mysteries about LLM's true nature remain unsolved, its applications are flourishing. During research, we discovered characteristics worth exploring in LLM-RAG system processes for AI speculative design SDGs roundtable simulations, including the influence of model types and data volumes on system and RAG applications. Future research could explore whether this architecture could use different LLM framework technologies like knowledge distillation and fine-tuning for broader exploration in related fields.

Framework concepts similar to SDGs, as guidelines leading humanity toward the future, may gradually constrain human thinking. AI speculative design offers opportunities to break these framework limitations, including applying multimodal AI to different system processes and application domains, creating opportunities to stimulate imagination and discussion spaces. Simultaneously, we could attempt to make LLM's black-box nature more transparent in AI speculative design processes, potentially extracting clues for world improvement from these world-mirroring models.

## 8. References

For the full list of references, please refer to the separate References file.

## 9. Appendix

For detailed supplementary materials, please refer to the separate Appendix files.

# Reference

Agrawal, R., Imieliński, T., & Swami, A. (1993). Mining association rules between sets of items in large databases. Proceedings of the 1993 ACM SIGMOD International Conference on Management of Data, 207–216.

Aher, G., Arriaga, R. I., & Kalai, A. T. (2022). Using large language models to simulate multiple humans and replicate human subject studies. In arXiv [cs.CL]. arXiv. https://openreview.net/pdf?id=eYlLlvzngu

Allen, C., Biddulph, A., Wiedmann, T., Pedercini, M., & Malekpour, S. (2024). Modelling six sustainable development transformations in Australia and their accelerators, impediments, enablers, and interlinkages. Nature Communications, 15(1), 594.

Anderson, C. (2008, January 30). TED's nonprofit transition. https://www.ted.com/talks/chris_anderson_ted_s_nonprofit_transition?subtitle=zh-tw

Anderson, C. (2016). *TED TALKS 說話的力量: 你可以用言語來改變自己，也改變世界: TED 唯一官方版演講指南* [*TED talks: The official TED guide to public speaking*] (T. Lin, Trans.). Dakuai Publishing. (Original work published 2016) *(In Chinese)*

Anderson, C. (2016, April 19). TED's secret to great public speaking. https://www.ted.com/talks/chris_anderson_ted_s_secret_to_great_public_speaking/transcript?referrer=playlist-before_public_speaking&autoplay=true

Anderson, C. (2024, April 8). Introducing TED's new tagline. https://blog.ted.com/ideas-change-everything/

Anderson, C. (n.d.). Chris Anderson. Retrieved June 16, 2024, from https://www.ted.com/speakers/chris_anderson_ted

Anesa, P. (2018). The Popularization of Environmental Rights in TED Talks. Pólemos, 12(1), 203–219.

Angheloiu, C., Sheldrick, L., Tennant, M., & Chaudhuri, G. (2020). Future Tense: Harnessing Design Futures Methods to Facilitate Young People's Exploration of Transformative Change for Sustainability. World Futures Review, 12(1), 104–122.

BBC News 中文. (2019, June 10). *人工智能 70 年：科幻和現實的交融*. BBC. https://www.bbc.com/zhongwen/trad/science-48380424

BBC News 中文. (2020, December 30). *年終回顧：2020 年牽動人心的十大國際新聞*. BBC. https://www.bbc.com/zhongwen/trad/world-55258244

Bergöö, M., Ebneter, L., Bader, C., Ott, C., & Breu, T. (2019). Switzerland Sustainable Development Solutions Network Switzerland.

Bertram, N., Dunkel, J., & Hermoso, R. (2023). I am all EARS: Using open data and knowledge graph embeddings for music recommendations. Expert Systems with Applications, 229, 120347.

Bizer, C. (n.d.). SemanticWebTools - W3C Wiki. Retrieved June 23, 2024, from https://www.w3.org/wiki/SemanticWebTools

Bostock, M. (n.d.). *D3: Data-Driven Documents*. Retrieved June 23, 2024, from https://d3js.org/

Certain expenses of the United Nations (Article 17, paragraph 2, of the Charter). (1967). International Law Reports, 34, 281–396.

Chen, H., & Luo, X. (2019). An automatic literature knowledge graph and reasoning network modeling framework based on ontology and natural language processing. Advanced Engineering Informatics, 42, 100959.

Chen, X., Jia, S., & Xiang, Y. (2020). A review: Knowledge reasoning over knowledge graph. Expert Systems with Applications, 141, 112948.

Chew, R., Bollenbacher, J., Wenger, M., Speer, J., & Kim, A. (2023). LLM-Assisted Content Analysis: Using Large Language Models to Support Deductive Coding. In arXiv [cs.CL]. arXiv. http://arxiv.org/abs/2306.14924

Choi, N. (2023, October 30). The architecture of today's LLM applications. The GitHub Blog. https://github.blog/2023-10-30-the-architecture-of-todays-llm-applications/

Chopra, M., & Mason, E. (2015). Millennium Development Goals: background. Archives of Disease in Childhood, 100 Suppl 1, S2-4.

Chopra, S., Clarke, R. E., Clear, A. K., Heitlinger, S., Dilaver, O., & Vasiliou, C. (2022). Negotiating sustainable futures in communities through participatory speculative design and experiments in living. Proceedings of the 2022 CHI Conference on Human Factors in Computing Systems, 1–17.

CKIP Lab 中文詞知識庫小組. (n.d.). Retrieved June 25, 2024, from https://ckip.iis.sinica.edu.tw/project/embedding

Communications materials. (2015, January 28). United Nations Sustainable Development; United Nations: Sustainable Development Goals. https://www.un.org/sustainabledevelopment/news/communications-material/

Deng, Y., Xia, M., Cao, M., & Ma, H. (2022). KGSMS: Knowledge Graph Sample based Multi-agent Simulation. 2022 IEEE 2nd International Conference on Electronic Technology, Communication and Information (ICETCI), 478–482.

Depoix, J. (n.d.). youtube-transcript-api: This is a python API which allows you to get the transcript/subtitles for a given YouTube video. It also works for automatically generated subtitles and it does not require an API key nor a headless browser, like other selenium based solutions do! Github. Retrieved June 3, 2024, from https://github.com/jdepoix/youtube-transcript-api

Design for Sustainability Studio. (n.d.). *Speculative design x SDGs*. Retrieved June 20, 2024, from https://www.designforsustainability.studio/type/speculative-design

Dunne A., 丹恩, Raby F., & 雷比. (2019). 推測設計: 設計, 想像與社會夢想. 何樵暐工作室出版.

Dunne, A. (2008). Hertzian Tales: Electronic Products, Aesthetic Experience, and Critical Design. MIT Press.

Egami, S., Ugai, T., Oono, M., Kitamura, K., & Fukuda, K. (2023). Synthesizing Event-Centric Knowledge Graphs of Daily Activities Using Virtual Space. IEEE Access, 11, 23857–23873.

Ehrlinger, L., & Wöß, W. (2016). Towards a definition of knowledge graphs. Semantics and Pragmatics.

Erdl, A. (2023, March 13). SustainGraph – A knowledge graph for the UN sustainability goals. Graph Database & Analytics; Neo4j. https://neo4j.com/developer-blog/sustaingraph-un-sustainability-goals/

Farnsworth W. (2023). The Socratic method: A practitioner's handbook. Mai Tian.

Fensel, D., Şimşek, U., Angele, K., Huaman, E., Kärle, E., Panasiuk, O., Toma, I., Umbrich, J., & Wahler, A. (2020). Introduction: What Is a Knowledge Graph? In D. Fensel, U. Şimşek, K. Angele, E. Huaman, E. Kärle, O. Panasiuk, I. Toma, J. Umbrich, & A. Wahler (Eds.), Knowledge Graphs: Methodology, Tools and Selected Use Cases (pp. 1–10). Springer International Publishing.

Forlano, L. E., & Halpern, M. K. (2023). Speculative Histories, Just Futures: From Counterfactual Artifacts to Counterfactual Actions. ACM Trans. Comput.-Hum. Interact., 30(2), 1–37.

Fotopoulou, E., Mandilara, I., Zafeiropoulos, A., & Papavassiliou, S. (2022, January 7). SustainGraph Ontology. https://netmode.gitlab.io/sustaingraph-ontology/

Fotopoulou, E., Mandilara, I., Zafeiropoulos, A., Laspidou, C., Adamos, G., Koundouri, P., & Papavassiliou, S. (2022). SustainGraph: A knowledge graph for tracking the progress and the interlinking among the sustainable development goals' targets. Frontiers of Environmental Science & Engineering in China, 10. https://doi.org/10.3389/fenvs.2022.1003599

Franz, M. (n.d.). Cytoscape.js. Retrieved June 23, 2024, from https://js.cytoscape.org/

Gao, Y., Xiong, Y., Gao, X., Jia, K., Pan, J., Bi, Y., Dai, Y., Sun, J., Wang, M., & Wang, H. (2023). Retrieval-Augmented Generation for Large Language Models: A Survey. In arXiv [cs.CL]. arXiv. http://arxiv.org/abs/2312.10997

Gautam, B., Ramos Terrades, O., Pujadas-Mora, J. M., & Valls, M. (2020). Knowledge graph based methods for record linkage. Pattern Recognition Letters, 136, 127–133.

Gemini. (n.d.-a). Google DeepMind. Retrieved June 24, 2024, from https://deepmind.google/technologies/gemini/

Gemini. (n.d.-b). Google for Developers. Retrieved June 24, 2024, from https://ai.google.dev/gemini-api/docs/models/gemini

Global sustainable development report (GSDR) 2019. (n.d.). Retrieved June 3, 2024, from https://sdgs.un.org/gsdr/gsdr2019

Google for Developers. (n.d.). *Embeddings*. Google. Retrieved June 25, 2024, from https://developers.google.com/machine-learning/crash-course/embeddings/video-lecture?hl=zh-tw

Google. (n.d.). *Search: list*. Google for Developers. Retrieved June 3, 2024, from https://developers.google.com/youtube/v3/docs/search/list?hl=zh-tw

Google. (n.d.-a). *How Google's Knowledge Graph works*. Retrieved June 22, 2024, from https://support.google.com/knowledgepanel/answer/9787176?hl=en

Gore, A. (2006, June 27). Averting the climate crisis. https://www.ted.com/talks/al_gore_averting_the_climate_crisis?referrer=playlist-the_first_ted_talks_ever

Gruber, T. R. (1995). Toward principles for the design of ontologies used for knowledge sharing? International Journal of Human-Computer Studies, 43(5), 907–928.

Guo, J., Mohanty, V., Piazentin Ono, J. H., Hao, H., Gou, L., & Ren, L. (2024). Investigating Interaction Modes and User Agency in Human-LLM Collaboration for Domain-Specific Data Analysis. Extended Abstracts of the 2024 CHI Conference on Human Factors in Computing Systems, 1–9.

Hassabis, D. (2024, May 14). Gemini breaks new ground with a faster model, longer context, AI agents and more. Google. https://blog.google/technology/ai/google-gemini-update-flash-ai-assistant-io-2024/

Hitzler, P., Krotzsch, M., & Rudolph, S. (2009). Foundations of Semantic Web Technologies. Taylor & Francis.

Hugging Face. (n.d.). *Building a RAG system with Gemma, MongoDB and open source models – Hugging Face open-source AI cookbook.* Retrieved June 25, 2024, from https://huggingface.co/learn/cookbook/rag_with_hugging_face_gemma_mongodb

IBM Technology [@IBMTechnology]. (2022, February 9). What is a Knowledge Graph? Youtube. https://www.youtube.com/watch?v=y7sXDpffzQQ

IBM. (2023, December 18). *What is embedding?* Retrieved from https://www.ibm.com/topics/embedding

International Institute for Sustainable Development. (n.d.-a). *ECOSOC, UNGA Presidents look to Summit of the Future to accelerate SDGs*. SDG Knowledge Hub. Retrieved June 11, 2024, from https://sdg.iisd.org/news/ecosoc-unga-presidents-look-to-summit-of-the-future-to-accelerate-sdgs/

International Institute for Sustainable Development. (n.d.-b). *Guest article: Harnessing scientific evidence and decision making to accelerate the SDGs*. SDG Knowledge Hub. Retrieved June 11, 2024, from https://sdg.iisd.org/commentary/guest-articles/harnessing-scientific-evidence-and-decision-making-to-accelerate-the-sdgs/

International Institute for Sustainable Development. (n.d.-c). *UN Collaborative seeks input on draft citizen data framework*. SDG Knowledge Hub. Retrieved June 10, 2024, from https://sdg.iisd.org/news/un-collaborative-seeks-input-on-draft-citizen-data-framework/

International Institute for Sustainable Development. (n.d.-d). *UN Statistical Commission adopts 36 changes to SDG indicators*. SDG Knowledge Hub. Retrieved June 10, 2024, from https://sdg.iisd.org/news/un-statistical-commission-adopts-36-changes-to-sdg-indicators/

International Institute for Sustainable Development. (n.d.-e). *Zero draft of HLPF outcome commits to accelerated action in years to 2030*. SDG Knowledge Hub. Retrieved June 11, 2024, from https://sdg.iisd.org/news/zero-draft-of-hlpf-outcome-commits-to-accelerated-action-in-years-to-2030/

James, C. (2021). Improving the Scopus and Aurora queries to identify research that supports the United Nations Sustainable Development Goals (SDGs) 2021 [Data set]. Mendeley. https://doi.org/10.17632/9SXDYKM8S4.4

James, C. (2022). Elsevier 2022 sustainable development goals (SDG) mapping [Data set]. Elsevier BV. https://doi.org/10.17632/6BJY52JKM9.1

James, C. (2023). Elsevier 2023 sustainable development goals (SDGs) mapping [Data set]. Elsevier BV. https://doi.org/10.17632/Y2ZYY9VWZY.1

Jana, S., Biswas, R., Pal, K., Biswas, S., & Roy, K. (2024). The evolution and impact of large language model systems: A comprehensive analysis.

Johri, P., Khatri, S. K., Al-Taani, A. T., Sabharwal, M., Suvanov, S., & Kumar, A. (2021). Natural Language Processing: History, Evolution, Application, and Future Work. Proceedings of 3rd International Conference on Computing Informatics and Networks, 365–375.

Kim, R. E. (2023). Augment the SDG indicator framework. Environmental Science & Policy, 142, 62–67.

Knez, T., & Žitnik, S. (2023). Event-Centric Temporal Knowledge Graph Construction: A Survey. Science in China, Series A: Mathematics, 11(23), 4852.

Kumar, S., Kumar, N., & Vivekadhish, S. (2016). Millennium Development Goals (MDGs) to Sustainable Development Goals (SDGs): Addressing Unfinished Agenda and Strengthening Sustainable Development and Partnership. Indian Journal of Community Medicine: Official Publication of Indian Association of Preventive & Social Medicine, 41(1), 1.

Lambert, N., Castricato, L., von Werra, L., & Havrilla, A. (2022). Illustrating Reinforcement Learning from Human Feedback (RLHF). Hugging Face Blog.

langchain-ai. (n.d.). *langchain: 🦜 🔗 Build context-aware reasoning applications*. GitHub. Retrieved June 25, 2024, from https://github.com/langchain-ai/langchain

Laskar, M. T. R., Fu, X.-Y., Chen, C., & Tn, S. B. (2023). Building Real-World Meeting Summarization Systems using Large Language Models: A Practical Perspective. In arXiv [cs.CL]. arXiv. http://arxiv.org/abs/2310.19233

Leal Filho, W., Dibbern, T., Pimenta Dinis, M. A., Coggo Cristofoletti, E., Mbah, M. F., Mishra, A., Clarke, A., Samuel, N., Castillo Apraiz, J., Rimi Abubakar, I., & Aina, Y. A. (2024). The added value of partnerships in implementing the UN sustainable development goals. Journal of Cleaner Production, 438, 140794.

Lee, H.-Y. [@HungyiLeeNTU]. (2016, November 26). ML Lecture 14: Unsupervised Learning - Word Embedding. Youtube. https://www.youtube.com/watch?v=X7PH3NuYW0Q

Li, Z., Shi, Y., Liu, Z., Yang, F., Payani, A., Liu, N., & Du, M. (2024). Quantifying Multilingual Performance of Large Language Models Across Languages. In arXiv [cs.CL]. arXiv. http://arxiv.org/abs/2404.11553

Lin, L., & Long, D. (2023). Generative AI Futures: A Speculative Design Exploration. Proceedings of the 15th Conference on Creativity and Cognition, 380–383.

Liu, J. (11 2022). LlamaIndex. https://doi.org/10.5281/zenodo.1234

LlamaIndex. (n.d.-a). *Building performant RAG applications for production*. Retrieved June 25, 2024, from https://docs.llamaindex.ai/en/stable/optimizing/production_rag/

LlamaIndex. (n.d.-b). *High-level concepts - LlamaIndex*. Retrieved June 25, 2024, from https://docs.llamaindex.ai/en/stable/getting_started/concepts/

LlamaIndex. (n.d.-c). *Vector store index - LlamaIndex*. Retrieved June 25, 2024, from https://docs.llamaindex.ai/en/stable/module_guides/indexing/vector_store_index/

Ma, P., Ding, R., Wang, S., Han, S., & Zhang, D. (2023). InsightPilot: An LLM-Empowered Automated Data Exploration System. In Y. Feng & E. Lefever (Eds.), Proceedings of the 2023 Conference on Empirical Methods in Natural Language Processing: System Demonstrations (pp. 346–352). Association for Computational Linguistics.

Mabuza, M. P. (2020). The UN Millennium Development Goals (MDGs) and Sustainable Development Goals (SDGs). In M. P. Mabuza (Ed.), Evaluating International Public Health Issues : Critical Reflections on Diseases and Disasters, Policies and Practices (pp. 77 – 103). Springer Singapore.

Madslien, J. (n.d.). Dotcom bubble burst: 10 years on. BBC.

Mair, S., Jones, A., Ward, J., Christie, I., Druckman, A., & Lyon, F. (2018). A Critical Review of the Role of Indicators in Implementing the Sustainable Development Goals. In W. Leal Filho (Ed.), Handbook of Sustainability Science and Research (pp. 41–56). Springer International Publishing.

Manning, C. D. (2022). Human language understanding & reasoning. Daedalus, 151(2), 127–138.

Martineau, K. (2023, August 22). What is retrieval-augmented generation (RAG)? IBM Research; IBM. https://research.ibm.com/blog/retrieval-augmented-generation-RAG

Mattiello, E. (2017). The popularisation of science via TED Talks. The International Journal of Bilingualism: Cross-Disciplinary, Cross-Linguistic Studies of Language Behavior, 11, 77–106.

Meskus, M., & Tikka, E. (2024). Speculative approaches in social science and design research: Methodological implications of working in 'the gap' of uncertainty. Qualitative Research: QR, 24(2), 209–228.

Meta AI. (2020, September 28). *Retrieval-augmented generation: Streamlining the creation of intelligent natural language processing models.* Retrieved June 25, 2024, from https://ai.meta.com/blog/retrieval-augmented-generation-streamlining-the-creation-of-intelligent-natural-language-processing-models/

Microsoft. (n.d.). *Welcome to GraphRAG*. Retrieved August 12, 2024, from https://microsoft.github.io/graphrag/

Mitrović, I. (2020, July 21). SpeculativeEdu. SpeculativeEdu. https://speculativeedu.eu/case-study-supervised-machine-learning-trainer-3607a/

Mitrović, I., Auger, J., Hanna, J., & Helgason, I. (2021). Beyond Speculative Design: Past - Present - Future. SpeculativeEdu.

Morris, J. J., & Alam, P. (2012). Value relevance and the dot-com bubble of the 1990s. The Quarterly Review of Economics and Finance: Journal of the Midwest Economics Association, 52(2), 243–255.

Nadkarni, P. M., Ohno-Machado, L., & Chapman, W. W. (2011). Natural language processing: an introduction. Journal of the American Medical Informatics Association: JAMIA, 18(5), 544–551.

Neo4j. (2020, May 16). *Neo4j graph database & analytics*. https://neo4j.com/

Neo4j. (2024, February 3). *NODES 2023 – Using LLMs to convert unstructured data to knowledge graphs* [Video]. YouTube. https://www.youtube.com/watch?v=qLdkRReMPvM

NetworkX Developers. (n.d.). *NetworkX — NetworkX documentation*. Retrieved June 23, 2024, from https://networkx.org/

Nilashi, M., Keng Boon, O., Tan, G., Lin, B., & Abumalloh, R. (2023). Critical data challenges in measuring the performance of sustainable development goals: Solutions and the role of big-data analytics. Harvard Data Science Review, 5(3). https://doi.org/10.1162/99608f92.545db2cf

Nilsson, M., Griggs, D., & Visbeck, M. (2016). Policy: Map the interactions between Sustainable Development Goals. Nature, 534(7607), 320–322.

Pan, S., Luo, L., Wang, Y., Chen, C., Wang, J., & Wu, X. (2023). Unifying Large Language Models and Knowledge Graphs: A Roadmap. In arXiv [cs.CL]. arXiv. http://arxiv.org/abs/2306.08302

Pichai, S. (2023, December 6). Introducing Gemini: our largest and most capable AI model. Google. https://blog.google/technology/ai/google-gemini-ai/

Pichai, S. (2024, February 15). Our next-generation model: Gemini 1.5. Google. https://blog.google/technology/ai/google-gemini-next-generation-model-february-2024/

Pinker) 史迪芬‧平克(steven. (2015). 語言本能: 探索人類語言進化的奧秘(最新中文修訂版). 商周出版.

Pradhan, P., Costa, L., Rybski, D., Lucht, W., & Kropp, J. P. (2017). A systematic study of sustainable development goal (SDG) interactions. Earth's Future, 5(11), 1169–1179.

*Prompting techniques.* (n.d.). Prompting Guide. Retrieved June 25, 2024, from https://www.promptingguide.ai/techniques

Purvis, B., Mao, Y., & Robinson, D. (2019). Three pillars of sustainability: in search of conceptual origins. Sustainability Science, 14(3), 681–695.

RAG is failing when the number of documents increase. (2024, January 5). OpenAI Developer Community. https://community.openai.com/t/rag-is-failing-when-the-number-of-documents-increase/578498

Reid, M., Savinov, N., Teplyashin, D., Lepikhin, D., Lillicrap, T., Alayrac, J.-B., Soricut, R., Lazaridou, A., Firat, O., Schrittwieser, J., Antonoglou, I., Anil, R., Borgeaud, S., Dai, A. M., Millican, K., Dyer, E., Glaese, M., Sottiaux, T., Lee, B., … Fernando, N. (2024). Gemini 1.5: Unlocking multimodal understanding across millions of tokens of context. ArXiv, abs/2403.05530. https://doi.org/10.48550/arXiv.2403.05530

Retrieval augmented generation (RAG). (n.d.). *Prompting Guide*. Retrieved June 25, 2024, from https://www.promptingguide.ai/techniques/rag

Rosenbaum, A., Soltan, S., & Hamza, W. (2023, January 20). Using large language models (LLMs) to synthesize training data. Amazon Science. https://www.amazon.science/blog/using-large-language-models-llms-to-synthesize-training-data

Rushmer, R. K., Hunter, D. J., & Steven, A. (2014). Using interactive workshops to prompt knowledge exchange: a realist evaluation of a knowledge to action initiative. Public Health, 128(6), 552–560.

Sachs, J. D., Lafortune, G., Fuller, G., & Drumm, E. (2023). Implementing the SDG stimulus. Sustainable development report 2023. Dublin University Press. https://doi.org/10.25546/102924

Servaes, J. (2017). Introduction: From MDGs to SDGs. In J. Servaes (Ed.), Sustainable Development Goals in the Asian Context (pp. 1–21). Springer Singapore.

Silva, M. C., Eugénio, P., Faria, D., & Pesquita, C. (2022). Ontologies and Knowledge Graphs in Oncology Research. Cancers, 14(8). https://doi.org/10.3390/cancers14081906

Simeone, L., Mantelli, R., & Adamo, A. (2022). Pushing divergence and promoting convergence in a speculative design process: Considerations on the role of AI as a co-creation partner. DRS Biennial Conference Series. https://doi.org/10.21606/drs.2022.197

Simsek, U., Angele, K., Kärle, E., Opdenplatz, J., Sommer, D., Umbrich, J., & Fensel, D. (2021). Knowledge graph lifecycle: Building and maintaining knowledge graphs. Second International Workshop on Knowledge Graph Construction.

Singhal, A. (2012, May 16). Introducing the Knowledge Graph: things, not strings. Google. https://blog.google/products/search/introducing-knowledge-graph-things-not/

Sugimoto, C. R., & Thelwall, M. (2013). Scholars on soap boxes: Science communication and dissemination inTEDvideos. Journal of the American Society for Information Science and Technology , 64(4), 663–674.

Sustar, H., Mladenović, M. N., & Givoni, M. (2020). The Landscape of Envisioning and Speculative Design Methods for Sustainable Mobility Futures. Sustainability: Science Practice and Policy, 12(6), 2447.

TED Localization Team. (2024, May 20). *Announcing AI-adapted multilingual TED Talks, unblocking language barriers*. TED Blog. https://blog.ted.com/announcing-ai-adapted-multilingual-ted-talks/

TED. (n.d.-a). *History of TED*. Retrieved June 14, 2024, from https://www.ted.com/about/our-organization/history-of-ted

TED. (n.d.-b). *Ways to get TED Talks*. Retrieved June 16, 2024, from https://www.ted.com/about/programs-initiatives/ted-talks/ways-to-get-ted-talks

TED. (n.d.-c). *Ideas change everything*. Retrieved June 16, 2024, from https://www.ted.com/about

TED. (n.d.-d). *TED Talks*. https://www.ted.com/about/programs-initiatives/ted-talks

TED. (n.d.-e). *Sustainability*. Retrieved June 18, 2024, from https://www.ted.com/topics/sustainability

The Economist. (2014, September 27). *The causes of a welcome trend*. https://www.economist.com/international/2014/09/27/the-causes-of-a-welcome-trend

The Knowledge Graph Conference. (2023, June 19). *KGC 2023 masterclass: Build a semantic layer in a knowledge graph — Stardog* [Video]. YouTube. https://www.youtube.com/watch?v=FMILiljigeE

The new Ted.com lets you dig deeper into ideas and see your influence on how they spread. (2014, March 4). *TED Blog*. https://blog.ted.com/introducing-new-ted/

Tosun, J., & Leininger, J. (2017). Governing the Interlinkages between the Sustainable Development Goals: Approaches to Attain Policy Integration. Global Challenges (Hoboken, NJ), 1(9), 1700036.

Transitioning from MDGs to the SDGs. (2016). UNDP ; https://digitallibrary.un.org/record/4026493/files/1387900EN.pdf

United Nations Development Programme (UNDP). (n.d.). Background on the goals. https://www.undp.org/sdg-accelerator/background-goals

United Nations Economic Commission for Europe. (n.d.). *Fundamental principles of official statistics*. Retrieved June 13, 2024, from https://unece.org/statistics/FPOS

United Nations Statistics Division. (n.d.-a). Events. Retrieved June 14, 2024, from https://unstats.un.org/sdgs/meetings/iaeg-sdgs-meeting-04

United Nations Statistics Division. (n.d.-b). IAEG-SDGs. Retrieved June 11, 2024, from https://unstats.un.org/sdgs/iaeg-sdgs/

United Nations Statistics Division. (n.d.-c). IAEG-SDGs — SDG Indicators. https://unstats.un.org/sdgs/iaeg-sdgs/tier-classification/

United Nations Statistics Division. (n.d.-d). SDG Indicators. Retrieved May 30, 2024, from https://unstats.un.org/sdgs/indicators/indicators-list/

United Nations. (2000). *United Nations Millennium Declaration (A/RES/55/2)*. https://www.un.org/en/development/desa/population/migration/generalassembly/docs/globalcompact/A_RES_55_2.pdf

United Nations. (2008). *Background: The Millennium Development Goals*. https://www.un.org/millenniumgoals/bkgd.shtml

United Nations. (2014). *Fundamental Principles of Official Statistics: Resolution adopted by the General Assembly (A/RES/68/261)*. https://digitallibrary.un.org/record/766579/files/A_RES_68_261-EN.pdf

United Nations. (2015). *Transforming our world: The 2030 Agenda for Sustainable Development (A/RES/70/1)*. https://digitallibrary.un.org/record/3923923/files/A_RES_70_1-EN.pdf

United Nations. (2016). *Report of the Inter-Agency and Expert Group on Sustainable Development Goal Indicators: Note by the Secretariat (E/CN.3/2016/2/Rev.1)*. https://digitallibrary.un.org/record/821651/files/E_CN.3_2016_2_Rev.1-EN.pdf

United Nations. (2017). *Report of the Inter-Agency and Expert Group on Sustainable Development Goal Indicators: Note by the Secretariat (E/CN.3/2018/2)*. https://digitallibrary.un.org/record/3845195/files/E_CN-3_2018_2-EN.pdf

United Nations. (2018). *Report of the Inter-Agency and Expert Group on Sustainable Development Goal Indicators: Note by the Secretariat (E/CN.3/2019/2)*. https://digitallibrary.un.org/record/1660362/files/E_CN-3_2019_2-EN.pdf

United Nations. (2018). *Transforming our world: The 2030 agenda for sustainable development*. United Nations. https://sdgs.un.org/2030agenda

United Nations. (2020). *Report of the Inter-Agency and Expert Group on Sustainable Development Goal Indicators: Note by the Secretariat (E/CN.3/2021/2)*. https://digitallibrary.un.org/record/3897450/files/E_CN.3_2021_2-EN.pdf

United Nations. (2021). *Report of the Inter-Agency and Expert Group on Sustainable Development Goal Indicators: Note by the Secretariat (E/CN.3/2022/2)*. https://digitallibrary.un.org/record/3954879/files/E_CN.3_2022_2-EN.pdf

United Nations. (2022). *Report of the Inter-Agency and Expert Group on Sustainable Development Goal Indicators: Note by the Secretariat (E/CN.3/2023/2)*. https://digitallibrary.un.org/record/4000329/files/E_CN.3_2023_2-EN.pdf

United Nations. (2023). *Report of the Inter-Agency and Expert Group on Sustainable Development Goal Indicators: Note by the Secretariat (E/CN.3/2024/4)*. https://digitallibrary.un.org/record/4033230/files/E_CN.3_2024_4-EN.pdf

United Nations. (n.d.). *About the SDG acceleration actions*. United Nations. Retrieved June 11, 2024, from https://sdgs.un.org/partnerships/action-networks/acceleration-actions/about

United Nations. (n.d.-a). International Day for Disaster Risk Reduction. https://www.un.org/en/observances/disaster-reduction-day

United Nations. (n.d.-b). Our Work | United Nations. Retrieved June 7, 2024, from https://www.un.org/en/our-work

United Nations. (n.d.-c). *The 17 goals*. United Nations. https://sdgs.un.org/goals

United Nations. (n.d.-d). *Mainstreaming the 2030 Agenda for Sustainable Development*. United Nations. https://sustainabledevelopment.un.org/unsystem/mainstreaming

Unlocking progress :MDG acceleration on the road to 2015 : lessons from the MDG Acceleration Framework pilot countries. (2010). UNDP,. http://digitallibrary.un.org/record/700583/files/Unlocking%2520Progress_MAF%2520Lessons%2520from%2520Pilot%2520Countries_7%2520October%25202010.PDF

Vandemoortele, J. (2018). From simple-minded MDGs to muddle-headed SDGs*. Development Studies Research, 5(1), 83–89.

Varshney, T. (2024, February 20). Build an LLM-powered data agent for data analysis. NVIDIA Technical Blog. https://developer.nvidia.com/blog/build-an-llm-powered-data-agent-for-data-analysis/

Vercel. (n.d.). Github. Retrieved June 5, 2024, from https://github.com/vercel

Vinuesa, R., Azizpour, H., Leite, I., Balaam, M., Dignum, V., Domisch, S., Felländer, A., Langhans, S. D., Tegmark, M., & Fuso Nerini, F. (2020). The role of artificial intelligence in achieving the Sustainable Development Goals. Nature Communications, 11(1), 233.

Wakkary, R., Odom, W., Hauser, S., Hertz, G., & Lin, H. (2015). Material speculation: actual artifacts for critical inquiry. Proceedings of The Fifth Decennial Aarhus Conference on Critical Alternatives, 97–108.

Wang, W. (2021). Do influential videos empower innovation ？ evidence from TED talks. http://2021.cswimworkshop.org/wp-content/uploads/2021/06/cswim2021_paper_85.pdf

Ways to get TED talks. (n.d.). Retrieved June 16, 2024, from https://www.ted.com/about/programs-initiatives/ted-talks/ways-to-get-ted-talks

What is D3? (n.d.). *D3.js*. Retrieved June 23, 2024, from https://d3js.org/what-is-d3

Why do we need to keep talks within 18 minutes? (n.d.). *TED Help Center*. Retrieved June 17, 2024, from https://help.ted.com/hc/en-us/articles/360038669354-Why-do-we-need-to-keep-talks-within-18-minutes

Window.localStorage. (n.d.). MDN Web Docs. Retrieved June 5, 2024, from https://developer.mozilla.org/zh-TW/docs/Web/API/Window/localStorage

World Wide Web Consortium (W3C). (n.d.). *OWL - Semantic Web Standards*. https://www.w3.org/OWL/

World Wide Web Consortium (W3C). (n.d.-a). *RDF - Semantic Web Standards*. Retrieved June 23, 2024, from https://www.w3.org/RDF/

World Wide Web Consortium (W3C). (n.d.-b). *RDF-based semantics*. Retrieved June 23, 2024, from https://www.w3.org/2007/OWL/wiki/RDF-Based_Semantics

Wu, H., Wang, Z., Wang, K., Omran, P. G., & Li, J. (2023). Rule learning over knowledge graphs: A review. TGDK, 1, 7:1-7:23.

Wu, X., Fu, B., Wang, S., Liu, Y., Yao, Y., Li, Y., Xu, Z., & Liu, J. (2023). Three main dimensions reflected by national SDG performance. Innovation (Cambridge (Mass.)), 4(6), 100507.

Xiao, Y.-H. (2021). *Zui xin AI ji shu: Zhi shi tu pu ji ji shu gai nian da cheng* [最新AI 技術：知識圖譜集技術概念大成] (First ed.). Shen Zhi Shu Wei.

Yılmaz, N., & Atay, G. (2023). Conceptual analysis of livable cities in the context of Ted Talks. Journal of Design for Resilience in Architecture and Planning, 4(2), 175–188.

YouTube. (n.d.). *python/search.py* [Source code]. GitHub. Retrieved June 3, 2024, from https://github.com/youtube/api-samples/blob/master/python/search.py

Yumnam, G., Gyanendra, Y., & Singh, C. I. (2024). A systematic bibliometric review of the global research dynamics of United Nations Sustainable Development Goals 2030. Sustainable Futures, 7, 100192.

Zafar, M. (2024, September 27). More data makes RAG applications, worse - wAI industries - medium. WAI Industries. https://medium.com/wai-industries/more-data-makes-rag-applications-worse-452bd2b98638

Zhang, Z., Han, X., Liu, Z., Jiang, X., Sun, M., & Liu, Q. (2019). ERNIE: Enhanced Language Representation with Informative Entities. In arXiv [cs.CL]. arXiv. http://arxiv.org/abs/1905.07129

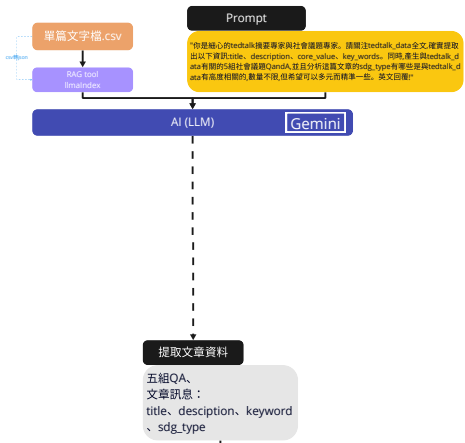
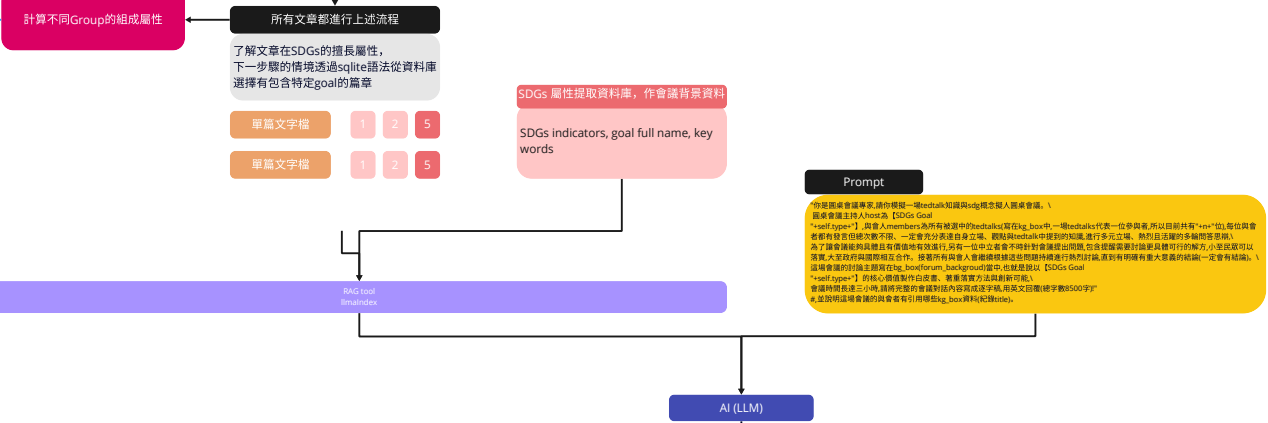
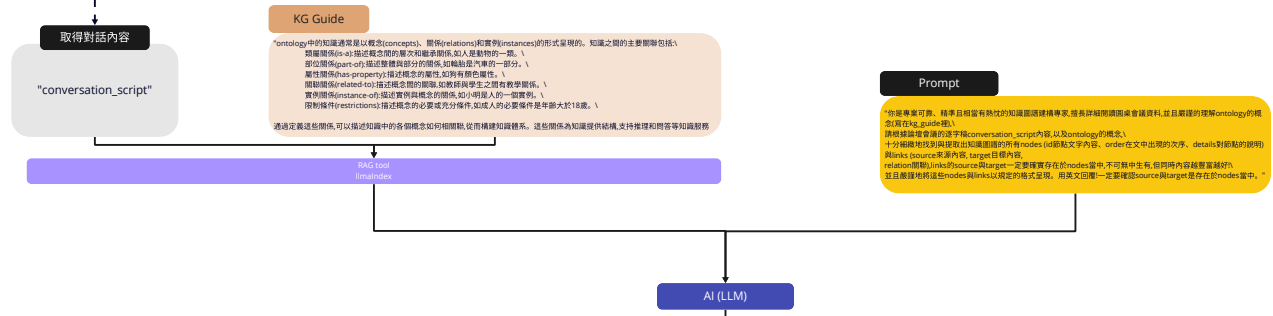
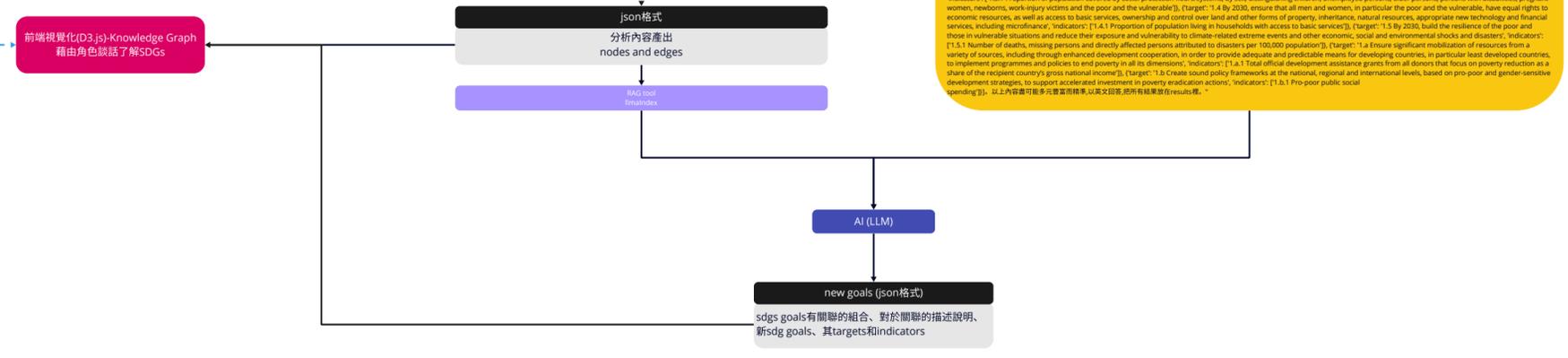

Appendix 2:Screenshots and Feature Description Summary of Knowledge Graphs from Roundtable Simulations for Each SDG Goal

| | Preliminary Dataset (269) | | | | | | | Formal Dataset (1127) | | | | | | |
|---|---|---|---|---|---|---|---|---|---|---|---|---|---|---|
| Target | Initial Node | Most Connected Node | Final Node | Color Variety | Arrow Direction Trend | Total Nodes | Total Links | Target | Initial Node | Most Connected Node | Final Node | Color Variety | Arrow Direction Trend | Total Nodes |
| Goal 1. No Poverty | Eradicating Poverty | Eradicating Poverty, Girls' Success and Empowerment | Ethical Implementation | 3 | Tending towards outward direction (complex and intricate) | 24 | 18 | No Poverty | No Poverty | Philanthropic Organizations | 3 | Pointing inwards | 14 | 16 |
| Goal 2. Zero Hunger | Zero Hunger | Zero Hunger, Sustainable Rice Farming | Target Interventions and Humanitarian Assistance | 3 | Pointing inwards | 16 | 15 | Zero Hunger | Zero Hunger | Cultural Food Traditions | 4 | Pointing outwards | 19 | 18 |
| Goal 3. Good Health & Well-Being | Goal 3: Good Health and Well-being | Goal 3: Good Health and Well-being | Environmental Issues | 3 | Pointing inwards | 11 | 10 | Goal 3: Good Health and Well-being | Goal 3: Good Health and Well-being | Stress Reduction Techniques | 4 | Pointing inwards | 15 | 14 |

| | Preliminary Dataset (269) | | | | | | Formal Dataset (1127) | | | | | |
|---|---|---|---|---|---|---|---|---|---|---|---|---|
| Target | Initial Node | Most Connected Node | Final Node | Color Variety | Arrow Direction Trend | Total Nodes | Total Links | Target | Initial Node | Most Connected Node | Final Node | Color Variety | Arrow Direction Trend | Total Nodes |
| Goal 4. Quality Education | SDG 4 - Quality Education | SDG 4 - Quality Education | Social-emotional learning | 3 | Pointing inwards | 16 | 15 | SDGs Goal4: Quality Education | White Paper | Transparency and Accountability in AI | 5 | Tending towards outward direction (complex and intricate) | 22 | 21 |
| Goal 5. Gender Equality | SDG Goal 5 | SDG Goal 5 | Data and Evidence | 3 | Pointing outwards | 16 | 16 | SDGs Goal 5 | Implementation Strategies | International Development | 3 | Pointing outwards | 15 | 14 |
| Goal 6. Clean Water & Sanitation | SDG 6: Clean Water and Sanitation | Sustainable Living | Decision-making Processes | 6 | Tending towards outward direction (complex and intricate) | 26 | 25 | SDG6 | White Paper | Roles and Responsibilities | 4 | Pointing outwards (multiple breakpoints/disruptions) | 13 | 7 |
| Goal 7. Affordable & Clean Energy | AI Impacts (kg_box) * kg_box in the prompt is a collective term for the TED Talks data. | AI Impacts (kg_box) * kg_box in the prompt is a collective term for the TED Talks data. | Explainable AI Solutions | 3 | Pointing outwards | 14 | 9 | Goal 7: Affordable and Clean Energy | Goal 7: Affordable and Clean Energy, Progress Measurement | Global Challenges | 6 | Pointing outwards (particularly evident later) | 31 | 30 |

| | Preliminary Dataset (269) | | | | | | | Formal Dataset (1127) | | | | | |
|---|---|---|---|---|---|---|---|---|---|---|---|---|---|
| Target | Initial Node | Most Connected Node | Final Node | Color Variety | Arrow Direction Trend | Total Nodes | Total Links | Target | Initial Node | Most Connected Node | Final Node | Color Variety | Arrow Direction Trend | Total Nodes |
| Goal 8. Decent Work & Economic Growth | Decent Work and Economic Growth | Decent Work and Economic Growth, Data Ownership, Boredom, AGI | Problem-Solving | 5 | Tending towards inward direction (complex and intricate) | 30 | 29 | Decent Work and Economic Growth | Decent Work and Economic Growth | Well-being and Sustainability | 4 | Pointing inwards | 18 | 17 |
| Goal 9. Industry, Innovation & Infrastructure | Goal 9 | Goal 9 | Inclusive Innovation Policies | 2 | Pointing inwards | 14 | 13 | Innovation | Innovation, Holistic Approach, Ethical Imperative | Environmental Stewardship | 3 | Pointing outwards | 17 | 15 |
| Goal 10. Reduced Inequalities | SDGs Goal 10: Reduced Inequalities | SDGs Goal 10: Reduced Inequalities | International Cooperation | 3 | Pointing inwards | 22 | 21 | Equitable Access to Quality Education | Individual Action | Philanthropy and Impact Investing | 3 | Pointing outwards | 12 | 6 |
| Goal 11. Sustainable Cities & Communities | SDGs Goal 11: Sustainable Cities and Communities | SDGs Goal 11: Sustainable Cities and Communities | Promoting Social Equality in Urban Development | 2 | Pointing inwards | 14 | 13 | Sustainable Cities and Communities | Sustainable Cities and Communities | Global Collaboration | 3 | Pointing inwards | 10 | 9 |

| | Preliminary Dataset (269) | | | | | | | Formal Dataset (1127) | | | | | | |
|---|---|---|---|---|---|---|---|---|---|---|---|---|---|---|
| Target | Initial Node | Most Connected Node | Final Node | Color Variety | Arrow Direction Trend | Total Nodes | Total Links | Target | Initial Node | Most Connected Node | Final Node | Color Variety | Arrow Direction Trend | Total Nodes |
| Goal 12. Responsible Consumption & Production | SDGs Goal 12 | SDGs Goal 12 | International Cooperation | 2 | Pointing inwards | 14 | 13 | SDG 12 | SDG 12 | Collaboration and Partnerships | 2 | Pointing outwards | 11 | 10 |
| Goal 13. Climate Action | SDGs Goal 13: Climate Action | SDGs Goal 13: Climate Action | Circular Food System | 3 | Pointing outwards | 15 | 14 | Climate Change | Climate Change | Vision | 4 | Pointing outwards | 40 | 39 |
| Goal 14. Life Below Water | SDG 14 - Life Below Water | SDG 14 - Life Below Water | Regenerative Ocean Farming | 5 | Pointing inwards | 16 | 16 | Goal 14: Life Below Water | Goal 14: Life Below Water | Harmony with the Ocean | 4 | Pointing outwards | 14 | 13 |
| Goal 15. Life on Land | SDG 15: Life on Land | SDG 15: Life on Land | Global collaboration | 2 | Pointing inwards | 20 | 19 | SDG 15 | SDG 15 | Education and Awareness | 2 | Pointing inwards | 9 | 8 |
| Goal 16. Peace, Justice & Strong Institutions | SDG 16 - Peace, Justice, and Strong Institutions | SDG 16 - Peace, Justice, and Strong Institutions | Citizen Diplomacy | 4 | Pointing inwards | 17 | 16 | SDG 16 | Disinformation | White Paper | 6 | Tending towards outward direction (complex and intricate) | 42 | 29 |

| | Preliminary Dataset (269) | | | | | | | Formal Dataset (1127) | | | | | |
|---|---|---|---|---|---|---|---|---|---|---|---|---|---|
| Target | Initial Node | Most Connected Node | Final Node | Color Variety | Arrow Direction Trend | Total Nodes | Total Links | Target | Initial Node | Most Connected Node | Final Node | Color Variety | Arrow Direction Trend | Total Nodes |
| Goal 17. Partnerships for the Goals | SDG 17 - Partnerships for the Goals | SDG 17 - Partnerships for the Goals, Online Communities | Global Collaboration on Mechanisms | 3 | Tending towards pointing inwards | 16 | 14 | Partnerships for the Goals | Partnerships for the Goals | Technology Serving Humanity | 7 | Tending towards inward direction (complex and intricate) | 41 | 37 |

**Note:** In the cell corresponding to the node with the most connections, the colored area indicates the initial node and will be a different node.

Appendix 3: Table of New SDGs Goal Composition and Descriptions

| Dataset | Goal | Sub-goal | Indicator | Source | Description |
|---|---|---|---|---|---|
| Preliminary Dataset | Goal 18: Inclusive Well-being | 18.1 By 2030, ensure that all individuals have equal opportunities to attain the highest possible level of health and well-being, regardless of their socioeconomic status, race, gender, or other characteristics. | 18.1.1 Proportion of the population reporting good health and well-being, disaggregated by socioeconomic status, race, gender, and other relevant characteristics., 18.1.2 Ratio of health expenditure coverage between the richest and poorest quintiles of the population. | Goal 3: Good Health and Well-being, SDGs  Goal 10: Reduced Inequalities | The knowledge graph highlights a relationship between Goal 3: Good Health and Well-being and SDGs Goal 10: Reduced Inequalities, suggesting that achieving good health and well-being is related to reducing inequalities. This connection is based on the presence of these goals as nodes in the graph and the existence of edges or paths connecting them. |
| Formal Dataset | Goal 18: Poverty Reduction through Technological Advancement | 18.1 By 2035, ensure universal access to affordable and relevant technology, particularly in impoverished communities, to bridge the digital divide and foster economic opportunities. | 18.1.1 Proportion of households in impoverished communities with access to affordable internet and digital devices., 18.1.2 Number of individuals in impoverished communities trained | Goal 1: No Poverty, Goal 9: Innovation | Innovation is presented as a key driver for eradicating poverty. |

| Dataset | Goal | Sub-goal | Indicator | Source | Description |
| --- | --- | --- | --- | --- | --- |
| | | | in digital literacy and technology-based skills. | | |
| | Goal 19: Climate Resilience for Vulnerable Communities | 19.1 By 2040, enhance the resilience of impoverished communities to climate-related extreme events and other environmental shocks. | 19.1.1 Number of climate-resilient infrastructure projects implemented in impoverished communities., 19.1.2 Proportion of households in impoverished communities with access to early warning systems for climate-related disasters. | Goal 1: No Poverty, Goal 13: Climate Change | Climate change is acknowledged as an interconnected issue with poverty, requiring investment in sustainable development initiatives to address both challenges. |
| 正式資料文本 | Goal 20: Inclusive and Equitable Development | 20.1 By 2045, eliminate all forms of discrimination and promote equal opportunities for all individuals, regardless of gender, race, ethnicity, religion, or socioeconomic status. | 20.1.1 Gender pay gap across different sectors and industries., 20.1.2 Proportion of leadership positions held by women and marginalized groups in various sectors. | Goal 5: Gender Equality, Goal 10: Reduced Inequalities | Implementation strategies for gender equality address intersectionality, which relates to social justice, a key aspect of reducing inequalities. |

| Dataset | Goal | Sub-goal | Indicator | Source | Description |
| --- | --- | --- | --- | --- | --- |
| | Goal 21: Global Collaboration for Water Security | 21.1 By 2050, establish a global alliance for water security, fostering collaboration among nations, organizations, and communities to address water challenges. | 21.1.1 Number of international agreements and partnerships focused on water security., 21.1.2 Amount of funding allocated to collaborative water management projects. | Goal 6: Clean Water and Sanitation, Goal 17: Partnerships for the Goals | Achieving SDG 6 requires collaboration among stakeholders, including policymakers, organizations, and individuals, highlighting the importance of partnerships. |
| | Goal 22: Inclusive Economic Empowerment | 22.1 By 2030, promote inclusive and sustainable economic growth, providing opportunities for decent work for all, particularly in impoverished communities. | 22.1.1 Employment rate in impoverished communities., 22.1.2 Average income of individuals in impoverished communities. | Goal 8: Decent Work and Economic Growth, Goal 1: No Poverty | Sustainable economic growth and decent work are crucial for poverty eradication., |

**Note**：Source Link: https://kg-web-4-0.vercel.app/new.html